\documentclass[acmsmall]{acmart}

\usepackage{lineno}   
\usepackage{booktabs} 
\usepackage{url}  

\usepackage{multirow}  
\usepackage{color}  
\newtheorem{strategy}{Strategy}    
\newtheorem{upper bound}{Upper bound}

\usepackage{diagbox}
\usepackage{amsmath}
\usepackage{arydshln} 
\usepackage{color} 
\usepackage{booktabs} 
\usepackage{multirow} 
\usepackage{makecell} 

\usepackage[linesnumbered,ruled]{algorithm2e} 

\SetAlFnt{\small}
\SetAlCapFnt{\small}
\SetAlCapNameFnt{\small}
\SetAlCapHSkip{0pt}
\IncMargin{-\parindent}

\usepackage{xspace}
\usepackage{multirow}
\usepackage{subfigure}

\setcopyright{rightsretained}
\acmJournal{JACM}

\begin{document}
	
\title{TaSPM: Targeted Sequential Pattern Mining}

\author{Gengsen Huang}
\affiliation{ 
	\institution{Jinan University}
	\city{Guangzhou}
	\country{China}
}
\email{hgengsen@gmail.com}

\author{Wensheng Gan}
\authornote{This is the corresponding author}
\affiliation{
	\institution{Jinan University; Pazhou Lab}
	\city{Guangzhou}
	\country{China}
}
\email{wsgan001@gmail.com}

\author{Philip S. Yu}
\affiliation{%
	\institution{University of Illinois at Chicago}
	\city{Chicago}
	\country{USA}
}
\email{psyu@uic.edu}


\begin{abstract}

Sequential pattern mining (SPM) is an important technique of pattern mining, which has many applications in reality. Although many efficient sequential pattern mining algorithms have been proposed, there are few studies can focus on target sequences. Targeted querying  sequential patterns can not only reduce the number of sequences generated by SPM, but also improve the efficiency of users in performing pattern analysis. The current algorithms available on targeted sequence querying are based on specific scenarios and cannot be generalized to other applications. In this paper, we formulate the problem of targeted sequential pattern mining and propose a generic framework namely TaSPM, based on the fast CM-SPAM algorithm. What's more, to improve the efficiency of TaSPM on large-scale datasets and multiple-items-based sequence datasets, we propose several pruning strategies to reduce meaningless operations in mining processes. Totally four pruning strategies are designed in TaSPM, and hence it can terminate unnecessary pattern extensions quickly and achieve better performance. Finally, we conduct extensive experiments on different datasets to compare the existing SPM algorithms with TaSPM. Experiments show that the novel targeted mining algorithm TaSPM can achieve faster running time and less memory consumption.

\end{abstract}


\authorsaddresses{\textbf{Authors' addresses}: 
Gengsen Huang, Jinan University, Guangzhou, China, hgengsen@gmail.com; Wensheng Gan, Jinan University, Guangzhou, China and also Pazhou Lab, Guangzhou 510335, China, wsgan001@gmail.com; Philip S. Yu, University of Illinois at Chicago, Chicago, USA, psyu@uic.edu
}


\renewcommand\shortauthors{W. Gan et al.}

\maketitle

\section{Introduction}

In the data mining area, frequent pattern mining (FPM) \cite{agrawal1994fast} is a well established topic, and it aims to discover all frequent and valuable patterns from the database provided by the user. As for how to specify whether a pattern is frequent or not, FPM is done by comparing the number of the pattern appearing in the database with the minimum support threshold (\textit{minsup}) specified by the user. For a frequent pattern, the items that appear in the pattern can be considered disordered. This means that $<$$A$, $B$$>$ and $<$$B$, $A$$>$ are the same, since FPM does not consider the order in which $A$ and $B$ appear in the transaction record. FPM is widely applied to various services in reality, and it can also be used to develop reasonable solutions and decisions based on different types of data. For example, by mining the data of a certain breakfast store, we can extract two frequent patterns $<$\textit{bread}$>$ and $<$\textit{bread}, \textit{milk}$>$. We can learn that most people make both purchases. Obviously, from these two frequent patterns, we can know that the main sales items of this breakfast store are bread and milk. The merchant can put bread and milk on top of the same shopping shelf, which makes more customers buy bread and milk at the same time. According to the actual survey, only those who have purchased bread will continually to buy milk, and if we only obtain this pattern $<$\textit{bread}, \textit{milk}$>$, we would not be able to know this relationship. The reason for this may be that bread is a staple food and milk is only a beverage. Therefore, in the case that customers basically buy bread, some of them will then buy milk. On the contrary, there will not be many customers who buy milk separately or continue to buy bread after buying milk, and we can know these situations from the infrequent pattern $<$\textit{milk}$>$. Based on the above analysis, the merchant can set a slight increase in the price of bread without increasing the price of the combination of bread and milk. Clearly, this decision can better promote sales. With this simple breakfast store case, we can clearly see the deficiency of FPM, which ignores the order of items appearing in the transaction records.

In order to solve the problems discussed above, sequential pattern mining (SPM) \cite{agrawal1995mining,fournier2017survey} was proposed. There is no doubt that SPM can find out more valuable and meaningful patterns than FPM. SPM takes into account the relationship between where each item appears in the transaction data, making $<$$A$, $B$$>$ and $<$$B$, $A$$>$ into two completely different patterns. There are a large number of related researchers have proposed a series of approaches and algorithms \cite{fournier2017survey,gan2019survey,mooney2013sequential} to improve efficiency. Besides, some researchers focus with specific research contexts in order to solve practical problems in different scenarios. For example, user behavior analysis in business \cite{srikant1996mining}, DNA analysis in biology \cite{floratou2011efficient}, and click-stream analysis on web pages \cite{fournier2012using}. Considering the format requirements of the discovered sequential patterns, there are also relevant studies introduced, such as closed sequential pattern \cite{gomariz2013clasp, wang2007frequent, yan2003clospan}, multi-dimensional sequential pattern \cite{pinto2001multi, songram2008closed}, and nonoverlapping sequential pattern \cite{wu2017nosep, wu2020netncsp}. Traditional SPM algorithms are based on the metric of frequency, and they aim to discover sequential patterns that satisfy the \textit{minsup}. Those mined sequential patterns can be considered as frequent sub-sequences of transaction records. The sequence in a transaction database can be classified into single-item-based sequence and multiple-items-based sequence depending on whether the itemset of the sequence contains more than one item. In practical applications, multiple-items-based sequences are often more valuable. For example, in a breakfast store, if a customer buys bread and milk simultaneously, the sequential pattern should be represented as $<$(\textit{bread}, \textit{milk})$>$, not $<$(\textit{bread}), (\textit{milk})$>$. The former means that the customer buys bread and milk at one time, while the latter means that the customer buys milk after buying bread. When comparing FPM and SPM, the sequential patterns mined by SPM will be much larger than those of FPM. This is because SPM takes into account the chronological order of each item. As we discussed above, $<$$A$, $B$$>$ and $<$$B$, $A$$>$ are the same pattern in FPM, while they are two different patterns in SPM.

Obviously, these algorithms will be very inefficient when dealing with large database or performing query functions. This is due to the large number of meaningless sequential patterns that are generated during the mining process. In the sequential pattern query function, there are top-$k$ sequential pattern querying \cite{fournier2013tks, tzvetkov2005tsp} and targeted sequential pattern querying \cite{chand2012target, chiang2003goal, chueh2010mining, zhang2021tusq}. The former aims at mining the top-$k$ most critical sequential patterns, while the latter mines the sequential patterns that users focus on. Targeted SPM aims to discover the patterns required by the user, thus avoiding the generation of irrelevant patterns. For example, in a supermarket, there will be a large number of transaction records. If we want to know which frequent sequential patterns are related to pencils, then other frequent sequential patterns should not be generated. In addition, web search is also a good example of targeted sequential pattern query. When the user enters keywords in the browser through the keyboard, the search engine will return to the user the relevant search records. Although some targeted SPM algorithms \cite{chand2012target, chiang2003goal, chueh2010mining} based on frequency have been proposed, they are incomplete or not general. These algorithms focus only on those sequential patterns where the last itemset is the query itemset. Such restrictions are clearly not suitable for web search and transaction log analysis. For example, if we want to query a page with a certain key verb, we may get very few results. This is because the verb does not usually appear at the end of the sentence. Furthermore, it is extremely unreasonable to focus on only one itemset. In transactional data, these algorithms are unable to provide relevant queries if we focus on the sequential patterns of users purchasing both books and pencils.

To address these issues, in this paper, we redefine the concept of target sequential pattern and then formulate the problem of targeted sequential pattern mining (TaSPM). We also introduce a post-processing technique and four strategies including unpromising transaction filter pruning strategy (UTFP), unpromising prefix item pruning strategy (UPIP), unpromising $S$-Extension item pruning strategy (USIP), and unpromising $I$-Extension item pruning strategy (UIIP). Based on these pruning strategies, we propose the TaSPM algorithm and its variants. The major contributions of our work in this paper are as follows:

\begin{itemize}
	\item Targeted mining makes more sense than mining all. In view of the problems of the current targeted pattern mining/searching, we redefine target sequential pattern and formalize the problem of targeted sequential pattern mining in this paper.
	
	\item  For a query sequence and the given database, we use a post-processing technique to save all eligible target sequential patterns. Furthermore, an efficient algorithm called TaSPM is proposed to find out a complete set of target sequential patterns. Based on the filtering approach, we propose UTFP to filter unrelated transactions.
	
	\item  Inspired by the idea of using location information to improve efficiency, we propose UPIP, USIP, and UIIP to reduce unnecessary operations in mining process. These strategies can well address the efficiency problem of TaSPM on large datasets, especially on the multiple-items-based sequence dataset.
	
	\item  The real and synthetic datasets are used in the experiments. Experiments show that different variants of TaSPM (adopts different strategies) can achieve different degrees of optimized performance compared to the baseline CM-SPAM algorithm.
\end{itemize}

The remainder part of this paper is organized as follows. In Section \ref{sec:relatedwork}, we briefly review and summarize the previous related work. We present the relevant definitions and formalize the problem of targeted sequential pattern mining in Section \ref{sec:preliminaries}. And then, we propose the TaSPM algorithm in Section \ref{sec:algorithm}. The experimental results are shown and discussed in Section \ref{sec:experiments}. Finally, in Section \ref{sec:conclusion}, we summarize and look forward to future work.

\section{Related Work}
\label{sec:relatedwork}

In this section, we discuss sequential pattern mining (SPM) in Section \ref{section:SPM} and targeted pattern querying (TPQ) in Section \ref{section:TPQ}.

\subsection{Sequential Pattern Mining}
\label{section:SPM}

Sequential pattern mining (SPM) was originally proposed by Agrawal and Srikant \cite{agrawal1995mining} and provides decisions for a large number of business activities and public services. Given a sequence database and a prdefined minimum support (\textit{minsup}), SPM will discover all sequential patterns that are satisfy \textit{minsup}. There are many algorithms proposed for SPM to discover a complete set of sequential patterns. Some of the better known algorithms are AprioriAll \cite{agrawal1994fast}, GSP \cite{srikant1996mining}, SPADE \cite{zaki2001spade}, PrefixSpan \cite{han2001prefixspan}, SPAM \cite{ayres2002sequential}, LAPIN \cite{yang2007lapin}, FAST \cite{salvemini2011fast}, CM-SPADE \cite{fournier2014fast}, and CM-SPAM \cite{fournier2014fast}. To the best of our knowledge, AprioriAll is the first algorithm for SPM and is based on Apriori property. To improve the efficiency of AprioriAll, its improved algorithm called GSP was proposed by Srikant and Agrawal. Both AprioriAll and GSP are based on the downward-closure property Apriori to prune search space. For a sequential pattern, if it is infrequent, then all its supersequence patterns are infrequent too. Therefore, if the support of a sequence is less than \textit{minsup}, then the extension operation or growth process associated with it can be terminated. This approach is used for almost all traditional SPM algorithms. GSP \cite{srikant1996mining} is a breadth-first search algorithm which first scans the database to get all frequent 1-sequences, and then uses these frequent sequences to obtain frequent 2-sequences. Immediately afterwards, it gets 3-sequences, 4-sequences, etc. It obtains ($k$+1)-sequences by continuously merging $k$-sequences. For better efficiency, a hash tree can be used to store the data, reducing the number of original data sequences that need to be checked for candidates. However, in any case, this process of merging generates a large number of candidates, making the GSP algorithm inefficient. Subsequently, the pattern-growth-based PrefixSpan \cite{han2001prefixspan} was proposed to solve the efficiency problem of SPM. PrefixSpan is the most popular SPM algorithm to discover sequential patterns. With the projection mechanism and two extension approaches, PrefixSpan can achieve far better efficiency than GSP. However, PrefixSpan faces the dilemma of excessive memory usage when dealing with large datasets. To address this issue, some  breadth-first-search-based algorithms, such as SPADE and SPAM, are designed with a better database representation to save memory. SPADE \cite{zaki2001spade} reduces the number of database scans by using ID-List, and SPAM \cite{ayres2002sequential} also reduces that by using Bitmap. Both vertical database representations can reduce memory consumption. To solve the problem on dense datasets, LAPIN \cite{yang2007lapin} focuses on the position of the last item, allowing the algorithms LAPIN-PrefixSpan \cite{yang2007lapin} and LAPIN-SPAM \cite{yang2005improved} to achieve excellent results on dense datasets. Inspired by SPAM, FAST has improved the data structure by introducing indexed sparse id-lists, which reduce memory consumption by quickly calculating the support of candidates in candidates generation. Besides, co-occurrence Map (CMAP) is used in CM-SPADE and CM-SPAM to further reduce meaningless operations during pattern growth. Compared to other algorithms, CM-SPADE and CM-SPAM can achieve an order of magnitude difference on certain datasets.

In addition to the traditional SPM algorithms discussed above, some other types of SPM algorithms have been proposed. Maximal sequential pattern mining (MSPM) and closed sequential pattern mining (CSPM) are two of these types of algorithms. Compared to MSPM, the patterns obtained from CSPM are lossless. For a sequence $s$, if there is no supersequence with the same support as it and containing it, then $s$ is a closed sequence. For MSPM, it does not require that the support of supersequence of $s$ must be the same. Therefore, the set of sequential patterns mined by MSPM can be considered as a subset of the set of sequential patterns mined by CSPM. MFSPAN \cite{guan2005mining}, DIMASP \cite{garcia2006new}, MaxSP \cite{fournier2013mining}, and VMSP \cite{fournier2014vmsp} are the main algorithms of MSPM. And CSPM algorithm has some famous algorithms, including CloSpan \cite{yan2003clospan}, BIDE \cite{wang2007frequent}, ClaSP \cite{gomariz2013clasp}, CM-ClaSP \cite{fournier2014fast}, and CloFAST \cite{fumarola2016clofast}, etc. In addition to constraint the format of sequential patterns to reduce the number of sequential patterns obtained by mining, there are some types of algorithms that discover the user's desired patterns over the set of all sequential patterns. Top-$k$ sequential pattern mining (TSPM) and targeted sequential pattern mining (TaSPM) are this type of algorithms, which discover the sequential patterns that meet the user's requirements. Compared to MSPM and CSPM, TSPM and TaSPM are more from the user's point of view. TSPM discovers the top-$k$ sequential patterns that are most valuable to the user, and its main algorithms are TSP \cite{tzvetkov2005tsp}, TKS \cite{fournier2013tks}, and SkOPUS \cite{petitjean2016skopus}. For the existing TaSPM algorithms \cite{chand2012target, chiang2003goal, chueh2010mining}, they all focus on whether the last itemset of the sequential pattern obtained by mining is the targeted itemset. Obviously, these SPM algorithms have some drawbacks. Generally speaking, these algorithms was proposed to solve the problem that traditional SPM would discover a large number of sequential patterns in large datasets or with too little \textit{minsup}. Too many sequential patterns are not conducive to user analysis and can be considered to be very redundant. In addition to these types of SPM, there are also sequential generator pattern mining \cite{fournier2014vgen, gao2008efficient}, non-overlapping SPM \cite{wu2017nosep, wu2020netncsp}, high-utility SPM \cite{gan2021utility,gan2021explainable}, and compressing SPM \cite{lam2014mining}. 

\subsection{Targeted Pattern Querying}
\label{section:TPQ} 

In order to reduce the number of patterns obtained by mining and improve the analysis efficiency, targeted queries play an important role in this process. The earliest technique that used targeted queries was the association mining technique. Kubat \textit{et al.} \cite{kubat2003itemset} designed Itemset-Tree to mine which rules with a user-specified targeted itemset and also discussed three query cases, the most useful of which is that the query is performed by entering the targeted itemset. In these queries, the system discovers all the rules with the targeted itemset in its antecedent and that satisfy the minimum support or the minimum confidence. In order to better improve the efficiency of Itemset-Tree, MEIT \cite{fournier2013meit} was proposed. MEIT is a better data structure that reduces unnecessary nodes of Itemset-Tree, thus improving the efficiency of targeted queries in itemset mining and association rule mining. In order to solve the problem that the Itemset-Tree cannot utilize Apriori property and many invalid operations in the mining process, Lewis \textit{et al.} \cite{lewis2019enhancing} designed a fast generation process to improve efficiency. In addition, query-constraint-based ARM (QARM) \cite{abeysinghe2017query} was introduced to analyze a wide variety of clinical datasets in the National Sleep Research Resource. Recently, Shabtay \textit{et al.} \cite{shabtay2021guided} proposed GFP-Growth to deal with the problem of multitude-targeted itemsets mining in big data. For utility mining \cite{gan2021survey}, TargetUM \cite{miao2021targetum} was the first algorithm that can find out the high-utility targeted itemsets from transaction database. Based on a tree-based data structure called lexicographic querying tree, TargetUM proposed three effective pruning strategies to achieve a better efficiency.

All of the above algorithms are focused on targeted itemset mining or targeted association rule mining, from transaction database. As mentioned before, sequence data is quite different from transaction data. In an earlier work, Chiang \textit{et al.} \cite{chiang2003goal} introduced the definition of Goal-oriented sequential pattern, i.e., a sequential pattern in which the last itemset is the targeted itemset. An inefficient multi-phase algorithm, as well as concepts such as loss, time windows, and normalization are also proposed. Chueh \cite{chueh2010mining} reversed and filtered the database and considered time-intervals to discover all target sequential patterns. By considering the old customer recall strategy and RFM model, Chand \textit{et al.} \cite{chand2012target} proposed a PrefixSpan-based algorithm to discover the target sequences that satisfy multiple constraints. These targeted SPM algorithms concerns only on the last itemset of the pattern, and are not hold in many applications. Unlike the previous algorithms for mining target frequent sequences, the TUSQ algorithm \cite{zhang2021tusq} is designed based on the utility-driven applications \cite{gan2021survey,gan2020proum}. Inspired by these research works, we aim to provide a generic definition of the problem of targeted sequential pattern mining and propose new efficient algorithms.

\section{Preliminaries and Problem Formulation}
\label{sec:preliminaries}

In this section, we define some fundamental concepts and notations in targeted pattern mining. Subsequently, we formalize the problem definition of targeted sequential pattern mining. 

In this paper, we use letters to denote the basic unit, and each item can be considered as a mapping of real items. Let $I$ = \{$i_{1}$, $i_{2}$, $\cdots$, $i_{N}$\} be an infinite set of distinct items occurring in a transaction database. An itemset can considered as a subset of $I$, denoted as $X$ ($X$ $\subseteq$ $I$). All items in an itemset will be sorted according to alphabetical order, and the alphabetical order is presented as $\succ_{lex}$. For an itemset $X$, its size is the number of items that appear in it, denoted as $\vert$$X$$\vert$. For a sequence $S$ $=$ $<$$X_{1}$, $X_{2}$, $\cdots$, $X_{m}$$>$, its size is the number of itemsets it contains and its length is the number of all items that appear in it. For another sequence $T$ = $<$$Y_{1}$, $Y_{2}$, $\cdots$, $Y_{n}$$>$, if $S$ is a sub-sequence of $T$ ($S$ $\subseteq$ $T$), then $\exists$ \(m\), \(1 \leq k_{1} < k_{2} < ... < k_{m} \leq n\) such that \(\forall 1 \leq v \leq m,X_{v} \subseteq Y_{k_{v}}\). 

For example, given two sequences $S_1$ = $<$($f$), ($f$, $g$), ($f$)$>$ and $S_2$ = $<$($f$, $h$), ($f$, $g$), ($f$)$>$, the size and length of $S_1$ are 3 and 4 respectively. In addition, $S_1$ is a sub-sequence of $S_2$, this is because $S_1$ $\subseteq$ $S_2$.

\begin{table}[!htbp]
	\centering
	\caption{Sequence database}
	\label{table1}
	\begin{tabular}{|c|c|}  
		\hline 
		\textbf{sid} & \textbf{Sequence} \\
		\hline  
		\(s_{1}\) & $<$($g$), ($a$, $d$)$>$ \\ 
		\hline
		\(s_{2}\) & $<$($g$), ($a$, $b$, $c$, $d$), ($b$), ($f$)$>$ \\  
		\hline  
		\(s_{3}\) & $<$($g$), ($a$, $b$, $c$, $d$), ($a$, $b$), ($e$)$>$ \\
		\hline  
		\(s_{4}\) & $<$($a$), ($b$, $c$, $d$, $e$), ($e$), ($f$)$>$ \\
		\hline
	\end{tabular}
\end{table}

\begin{definition}[Sequence database]
	\rm For a sequence database $\mathcal{D}$, it is a set of tuples $<$\textit{sid}, $S$$>$ in which \textit{sid} denotes the identifier of the sequence $S$. The size of this set is also the size of the database $\mathcal{D}$, denoted as $\vert$$\mathcal{D}$$\vert$.
\end{definition}

To better illustrate the relevant definitions and concepts that follow, the sequence database we use as a running example is shown in Table \ref{table1}. We can see that this example database has five sequences, and each of them has a corresponding unique identifier. 

\begin{definition}[Support and minimum support]
	\rm For an item $e$ or an itemset $I$, its support can be denoted as \textit{sup}($e$ / $I$), defined as \textit{sup}($e$ / $I$) = $\vert${$S$$\vert$$e$ / $I$ $\in$ $S$ $\land$ $S$ $\in$ $\mathcal{D}$}$\vert$. Similarly, the support value of a sequence $S$ can be denoted as \textit{sup}($S$), defined as \textit{sup}($S$) = $\vert${$T$$\vert$$S$ $\subseteq$ $T$ $\land$ $T$ $\in$ $\mathcal{D}$}$\vert$. The minimum support is defined in advance by the user and is denoted as \textit{minsup}. For an item $e$ or a sequence $S$, if it satisfies \textit{sup}($e$ / $S$) $\ge$ \textit{minsup}, it is called a frequent item or sequential pattern. 
\end{definition}

For example in Table \ref{table1}, if \textit{minsup} is set to 2, then all items are frequent items. And $<$($g$), ($a$)$>$ is a frequent sequential pattern, because \textit{sup}($<$($g$), ($a$)$>$) = 3 $\ge$ 2.

\begin{definition}[Target sequential pattern and targeted sequential pattern mining]
	\rm Given a query sequence, called \textit{qs}, the target sequential pattern \textit{ts} of \textit{qs} will satisfy \textit{qs} $\subseteq$ \textit{ts} and \textit{sup}(\textit{ts}) $\ge$ \textit{minsup}, i.e. \textit{qs} is the frequent sub-sequence of \textit{ts}. The goal of targeted sequential pattern mining is to find out a complete set of target sequential patterns with the query sequence \textit{qs}. 
\end{definition}

For example in Table \ref{table1}, given a query sequence \textit{qs} = $<$($a$), ($b$)$>$ and \textit{minsup} = 2. \textit{ts}$_1$ = $<$($g$), ($a$), ($b$)$>$ is one of the target sequential patterns of \textit{qs}, while \textit{ts}$_2$ = $<$($g$), ($a$), ($b$), ($e$)$>$ is not, because \textit{sup}(\textit{ts}$_2$) = 1 $\textless$ \textit{minsup}. Based on the above concepts, a generic problem of targeted sequential pattern mining (TaSPM) can be formulated below.

\textbf{Problem statement}: Given a sequence database $\mathcal{D}$, a query sequence \textit{qs}, and a predefined minimum support threshold \textit{minsup}, the goal of TaSPM is to find out the complete set of target sequential patterns \textit{ts} of \textit{qs}, and they satisfy \textit{qs} $\subseteq$ \textit{ts} and \textit{sup}(\textit{ts}) $\ge$ \textit{minsup}.

\section{The TaSPM Algorithm}
\label{sec:algorithm}

In this section, we first present some important definitions and several pruning strategies. Subsequently,, we propose our efficient algorithm, called TaSPM, which is based on CM-SAPM \cite{fournier2014fast}.

\subsection{Definitions and pruning strategies}

\begin{definition}[Extension]
	\rm Extension is an important operation of the pattern-growth-based mining algorithm. There are two types of extensions used in the mining process, one of which is $S$-Extension and the other is $I$-Extension. The $S$-Extension adds a new itemset to the end of the current sequence, and that new itemset contains a new item. As for $I$-Extension, it adds a new item to the last itemset of the current sequence. Obviously, with these two extensions, the sequence will grow.
\end{definition}

\begin{definition}[Post-processing technique]
	\rm As we discussed above, there is no algorithm for the new definition of targeted sequential pattern mining. If we use a traditional sequential pattern mining algorithm to perform this task, then we need a post-processing technique. This technique works when saving a sequential pattern that mined by SPM. It scans the sequential pattern generated and determines if the sequential pattern contains the query sequence. If a pattern contains the query sequence, it will be saved, otherwise it will be filtered.
\end{definition}

\begin{definition}[Itemset match position and item match position]
	\rm In this paper, we specify that the corresponding item at the matching position of the query sequence is called current query item, denoted as \textit{qi}. During the pattern growth, we use two flags to record the current position match on the \textit{qi}. The first flag is called an itemset match position, denoted as \textit{IMatch}. \textit{IMatch} is used to record the position of the itemset currently matched to the query sequence. The second flag is called item match position and is denoted as \textit{IIMatch}, which is used to match the position of the item in itemset currently matched to the query sequence. 
\end{definition}

For example, given a query sequence $<$($g$), ($a$, $d$)$>$ and a current pattern $<$($g$), ($b$)$>$, we can know that the matching sequence is $<$($g$)$>$ and \textit{qi} is $a$. Therefore, \textit{IMatch} is equal to 1 and \textit{IIMatch} is equal to 0.

\begin{definition}[Bitmap representation of a sequence]
	\rm For a sequence $s$, if we represent it using a bitmap representation, it will be [$S_{11}$ $S_{12}$ ..., $S_{21}$ ..., ...].  In a sequence database, we use $S_{ij}$ to denote the position of $j$-th itemset in the $i$-th sequence. If the sequence $s$ can be matched by $S_{i}$ and the matching position is $j$, then $S_{ij}$ is equal to 1.
\end{definition}

For example in Table \ref{table1}, given a sequence \textit{s} = $<$($a$), ($b$)$>$, if we use a bitmap to represent this query sequence, then it would be [0 0, 0 0 1 0, 0 0 1 0, 0 1 0 0]. This is because $S_{2}$, $S_{3}$, and $S_{4}$ can all match $s$ and the positions are 2, 3, and 2 respectively.

\textbf{Question 1. When does it mean that the query sequence has been matched?}  We initialize \textit{IMatch} and \textit{IIMatch} to 0 when the sequence is $<$$>$. As the sequence grows, these two flags are updated. When \textit{IMatch} is equal to the size of the query sequence, it means that the query sequence is matched exactly. After that we can stop updating \textit{IMatch} and \textit{IIMatch} and start saving the target sequential patterns as well.

\textbf{Question 2. When will \textit{IMatch} and \textit{IIMatch} be updated and not be updated?}  We can decide whether to update \textit{IMatch} and \textit{IIMatch} based on whether the extended item of the current sequence matches to \textit{qi}. Initially, both \textit{IMatch} and \textit{IIMatch} will be initialized to 0.

When the current sequence is nothing, i.e. current sequence is $<$$>$ and \textit{qi} is the first item of query sequence. For an extended item $e$, if it does not match \textit{qi}, then both \textit{IMatch} and \textit{IIMatch} remains 0. If item $e$ matches \textit{qi}, then \textit{IIMatch} is updated to 1. Subsequently, if \textit{IIMatch} equals the size of the first itemset in query sequence, then \textit{IMatch} is updated to \textit{IIMatch} + 1 and \textit{IIMatch} is updated to 0. 
	
As the sequence grows, each extended item is compared with \textit{qi} of the query sequence. If the extended item $e$ is the same as the \textit{qi}, then \textit{IIMatch} is updated to \textit{IIMatch} + 1. After \textit{IIMatch} is updated, if \textit{IIMatch} satisfies \textit{IIMatch} equals the size of the \textit{IMatch}-th itemset of the query sequence, then \textit{IMatch} is updated to \textit{IMatch} + 1 and \textit{IIMatch} is updated to 0. 
	
There is another situation that  \textit{IIMatch} needs to be updated. When \textit{IIMatch} is not 0, but the current sequence performs $S$-Extension, this indicates that we need to reset the matching position, i.e., update \textit{IIMatch} to 0. If the extended item matches the item \textit{qi} after updating the position of the query sequence, then update \textit{IIMatch} to 1 and decide whether we need to continue updating \textit{IMatch} and \textit{IIMatch}.
	
As for the sequence performs $I$-Extension, if the extended item $e$ is less than \textit{qi}, we will not update \textit{IMatch} and \textit{IIMatch}. If the extended item $e$ and \textit{qi} are the same, then we update \textit{IIMatch}, and then repeat the above judgment process to update \textit{IMatch}. If the item $e$ is greater than \textit{qi}, the \textit{IIMatch} has to be updated to 0 since all the items in the itemset are sorted according to alphabetical order. In this case, the sequence continues to perform $I$-Extension is unable to match the itemset located at position \textit{IMatch} of the query sequence.
	
In addition, consider this case, where the current sequence is $<$($a$), ($b$)$>$ and the query sequence is $<$($a$), ($b$, $c$), ($e$)$>$. When we add item $c$ to current sequence through $I$-Extension, the sequence becomes $<$($a$), ($b$, $c$)$>$, and it can continually performs $I$-Extension. Since the extended item is $c$ is consistent with the item to be queried, then \textit{IIMatch} is updated and causes \textit{IMatch} to be updated. Obviously, if we continue to perform $I$-Extension by using item $e$, then it will cause \textit{IIMatch} to be updated. In this way, the matching position is wrong. It is because $<$($a$), ($b$, $c$, $e$)$>$ and $<$($a$), ($b$, $c$), ($e$)$>$ are not the same. The former cannot match the query sequence, while the latter can. To solve this problem, we use the flag \textit{NotUpdate} to determine if updates are not performed on \textit{IMatch} and \textit{IIMatch}. If \textit{NotUpdate} is true, we will not update the \textit{IMatch} and \textit{IIMatch}. When the growth sequence can still performs $I$-Extension after \textit{IMatch} is updated, \textit{NotUpdate} is updated to true. When the sequence performs $S$-Extension, the \textit{NotUpdate} is updated to false. There is also another case and we will not update \textit{IMatch} and \textit{IIMatch}, i.e., \textit{IMatch} is equal to the size of the query sequence, which means that the query sequence has been completely matched.

\begin{strategy}[Unpromising transaction filter pruning strategy, UTFP]
	\rm For a query sequence \textit{qs}, if a transaction record in the database $\mathcal{D}$ does not contain \textit{qs}, the transaction record is filtered. It is quite obvious that transaction records that do not contain \textit{qs} do not generate the target sequential patterns. By removing these irrelevant transaction records, TaSPM not only reduces the memory used to store these records, but also improves efficiency. If the size of database $\mathcal{D}$ is smaller than \textit{minsup} after the database $\mathcal{D}$ is filtered, it means that no frequent target sequential patterns can be mined.
\end{strategy}

For example in Table \ref{table1}, given a query sequence \textit{qs} = $<$($a$), ($f$)$>$, if we use a bitmap to represent this query sequence, then it would be [0 0, 0 0 0 1, 0 0 0 0, 0 0 0 1]. Obviously, we should filter the sequences $s_1$ and $s_3$ to obtain a more concise representation, i.e., [0 0 0 1, 0 0 0 1]. If we do not use unpromising transaction filter pruning strategy (UTFP) to filter these two sequences, when \textit{minsup} = 2 and the current sequence is $<$($g$)$>$, the support of $<$($g$)$>$ will be 3 and will continue to grow recursively. In this way, invalid operations become more numerous, because the support of the target sequential pattern with $g$ is equal to 1 at the highest. As another example, if the query pattern is $<$($a$), ($e$), ($f$)$>$, after the database $\mathcal{D}$ is filtered, then $\mathcal{D}$ contains only $s_4$. In this case, $\vert \mathcal{D} \vert$ is less than \textit{minsup}, then no frequent target sequential patterns will be generated.

\begin{strategy}[Unpromising prefix item pruning strategy, UPIP]
	\rm In the initial stage of the most sequential pattern mining algorithms, recursive growth is generally performed using frequent items. For a prefix item $e$, we compare the bitmap of $e$ with the bitmap of \textit{qi} and count the number of sequences where the position of \textit{qi} appears after the frequent item $e$. This number is denoted as \textit{fn}. In this case, \textit{qi} is also the first item in the query sequence. If item $e$ is not \textit{qi} and \textit{fn} is less than \textit{minsup}, we do not use item $e$ to continue to grow pattern. This is because the maximum support of any target sequential pattern beginning with $e$ will be less than \textit{minsup}, and we can safely assume that $e$ is not a beginning item to grow patterns.
\end{strategy}

For example in Table \ref{table1}, given a query sequence \textit{qs} = $<$($a$), ($b$)$>$ and \textit{minsup} is set to 2. We can know that item $f$ is a frequent item because \textit{sup}($f$) = 2 $\ge$ \textit{minsup}. Now let's do a bitmap comparison using UPIP. For this query sequence \textit{qs}, \textit{qi} is $a$, and its bitmap is [0 1 0 0, 0 1 0 0, 1 0 0 0]. For frequent item $f$, its bitmap is [0 0 0 1, 0 0 0 0, 0 0 0 1]. Obviously, according to UIUP, \textit{fn} will be equal to 0. Then we will not use the frequent item $f$ to grow pattern. In fact, item $f$ can be used as a part of the target sequential pattern, such as $<$($a$), ($b$), ($f$)$>$, \textit{sup}($<$($a$), ($b$), ($f$)$>$) = 2.

\begin{strategy}[Unpromising $S$-Extension item pruning strategy, USIP]
	\rm We propose USIP to decide which sequential patterns can continue to grow pattern recursively after being extended by $S$-Extension. We compare the bitmap of extended item $e$ with the bitmap of \textit{qi} and count the number of sequences where the position of \textit{qi} appears after the position of $e$. This number is denoted as \textit{fn}. If \textit{fn} is less than \textit{minsup}, we do not use item $e$ as an extended item to grow pattern. This is because the maximum support of the target sequential pattern with the extended sequential pattern as a prefix cannot be greater than \textit{minsup}. If the extended item $e$ matches \textit{qi}, then \textit{fn} will be changed to the number of sequences where the position of \textit{qi} is the same as or after the position of \textit{e}.
\end{strategy}

For example in Table \ref{table1}, given a query sequence \textit{qs} = $<$($a$), ($b$)$>$, a current sequence $s$ = $<$($g$)$>$, and \textit{minsup} = 2. Obviously, item $b$ is an extended item that can be extend by $S$-Extension, and extended sequence $s\prime$ = $<$($g$), ($b$)$>$ is a frequent pattern. In this case, \textit{qi} is item $a$. The bitmap of $s\prime$ is [0 1 1 0, 0 1 1 0, 0 0 0 0] and the bitmap of \textit{qi} is [0 1 0 0, 0 1 1 0, 1 0 0 0]. Therefore, \textit{fn} is equal to 1. $s\prime$ will not used to grow pattern. In fact, $s\prime$ can continue to be extended, but the sequential pattern obtained by its extension will not belong to the target sequential pattern of the query sequence \textit{qs}. If the extended item is $a$, then the bitmap of $s\prime$ will be [0 1 0 0, 0 1 1 0, 0 0 0 0] and \textit{fn} will be equal to 2 $\ge$ \textit{minsup}. Therefore, $<$($g$), ($a$)$>$ can continue to grow pattern. Finally, we can obtain a sequence $t$ = $<$($g$), ($a$), ($b$)$>$ that is a target sequential pattern of query sequence \textit{qs}.

\begin{strategy}[Unpromising $I$-Extension item pruning strategy, UIIP]
	\rm It is sufficient to use only USIP on a single-item-based sequence dataset, because the sequence can only perform $S$-Extension to grow pattern. However, on multiple-items-based sequence datasets, sequences can perform $I$-Extension, which requires UIIP to reduce meaningless extension operations. We propose UIIP to decide which sequential patterns can continue to grow pattern recursively after being extended by $I$-Extension. As we discussed above in Question 2, we do not need to update \textit{IMatch} and \textit{IIMatch} if the extended item is less than \textit{qi}. Otherwise, we need to update \textit{IIMatch} and \textit{IMatch}. If the match position is updated, \textit{qi} may also have been changed.
	
	For an extended item, there are two situations that we need to consider. The first one is item $e$ is less than or equal to \textit{qi}. In this case, we use \textit{fn} to count the number of sequences where the position of \textit{qi} is equal to or greater than the position of \textit{e}. The second case is when the item $e$ is greater than \textit{qi}, and \textit{IIMatch} will be updated to 0. we use \textit{fn} to count the number of sequences where the position of \textit{qi} greater than the position of \textit{e}. For these two cases, if \textit{fn} is less than \textit{minsup}, we do not use item $e$ as an extended item to grow pattern.
\end{strategy}

For example in Table \ref{table1}, given a query sequence \textit{qs} = $<$($g$), ($a$, $c$), ($b$)$>$, a current sequence $s$ = $<$($g$), ($a$)$>$, and \textit{minsup} = 2. Obviously, items $b$, $c$, and $d$ are extended items that can be extended by $I$-Extension. For item $b$, it is smaller than \textit{qi} and \textit{qi} will not be updated. The bitmap of extended sequence $s\prime$ is [0 1 0 0, 0 1 0 0] and the bitmap of \textit{qi} is [0 1 0 0, 0 1 0 0]. In this case, \textit{fn} is equal to 2 and $s\prime$ = $<$($g$), ($a$, $b$)$>$ can continually to grow pattern. For item $c$, it matches \textit{qi} and \textit{qi} will be updated to $b$. The bitmap of extended sequence $s\prime$ $=$ $<$($g$), ($a$, $c$)$>$ is [0 1 0 0, 0 1 0 0] and the bitmap of \textit{qi} is [0 0 1 0, 0 0 1 0]. In this case, \textit{fn} is equal to 2 and $s\prime$ can continually to grow pattern. As for the last item $d$ which is greater than \textit{qi}, and \textit{qi} will be updated to $a$. The bitmap of extended sequence $s\prime$ = $<$($g$), ($a$, $d$)$>$ is [0 1 0 0, 0 1 0 0] and the bitmap of \textit{qi} is [0 0 0 0, 0 0 1 0]. Therefore, \textit{fn} is equal to 1 and $s\prime$ will not be used to grow pattern.

\subsection{ Proposed TaSPM algorithm}

TaSPM utilizes a bitmap representation for all items and sequences, which is similar to CM-SPAM \cite{fournier2014fast}. Based on four novel pruning strategies, we propose the TaSPM algorithm for targeted discovering sequential patterns. TaSPM is a depth-first-search algorithm and the main procedure is shown in Algorithm \ref{AL:TaSPM}.

TaSPM first scans database $\mathcal{D}$ to filter the transaction record which does not contain the query sequence (Line 1). After the database is filtered, TaSPM scans the database $\mathcal{D}$ again to get all frequent items (denoted as \textit{F1}), removes all infrequent items from $\mathcal{D}$, creates the bitmap for all frequent items, and builds the \textit{CMAP} for co-occurrence pruning (Line 2). TaSPM initializes the relevant variables \textit{qi}, \textit{IMatch}, and \textit{IIMatch} for the next operations (Line 3). For all frequent item in \textit{F1}, TaSPM will get \textit{fn} and using UPIP to decide whether this prefix pattern can continually to grow (Lines 4-8). If a frequent prefix pattern satisfies the \textit{minsup}, TaSPM will use function \textit{UpdateMatch} to update the matching position (Line 9). The details of function \textit{UpdateMatch} can be referred to the above discussion, and do not describe here. At last, TaSPM will perform the recursive procedure to grow pattern (Line 10).

\begin{algorithm}[!htbp]
	\caption{The TaSPM algorithm}
	\label{AL:TaSPM}
	\LinesNumbered
	\KwIn{$\mathcal{D}$: a sequence database; \textit{qs}: a query sequence provided by user; \textit{minsup}: a minimum support defined by user.} 
	\KwOut{all target sequential patterns of query sequence \textit{qs}.}
	filter sequences that do not contain \textit{qs} in $\mathcal{D}$; \qquad(\textbf{UTFP Strategy}) \\
	scans the $\mathcal{D}$ again to 
	1) get all frequent items (\textit{F1}); 
	2) remove all infrequent items from $\mathcal{D}$;
	3) create the bitmap for all frequent items;
	4) build the \textit{CMAP}\;
	initializes \textit{qi} $\leftarrow$ the first item in \textit{qs}, \textit{IMatch} $\leftarrow$ 0, \textit{IIMatch} $\leftarrow$ 0\;
	\For{$<$$f$$>$ $\in$ \textit{F1}}{
		initialize \textit{fn} $\leftarrow$ the result of comparing the bitmap of $f$ with the bitmap of \textit{qi}\;
		\If{\textit{fn} $\textless$ \textit{minsup}}{
			continue; \qquad(\textbf{UPIP Strategy})
		}
		\textit{IMatch}, \textit{IIMatch} = call \textbf{UpdateMatch}()\;
		call \textbf{SEARCH}($<$$f$$>$, \textit{F1}, \{$e$ $\in$ \textit{F1} $\vert$ $e$ $\succ_{lex}$  $f$\}, \textit{qs}, \textit{minsup}, \textit{IMatch}, \textit{IIMatch}).
	}
\end{algorithm}

\begin{algorithm}[htbp]
	\caption{The SEARCH procedure}
	\label{AL:SEARCH}
	\LinesNumbered	
	\KwIn{$s$: the current sequence; \textit{SE}: a set containing all items can perform $S$-Extention; \textit{IE}: a set containing all items can perform $I$-Extention; \textit{qs}: a query sequence provided by user; \textit{minsup}: a minimum support defined by user; \textit{IMatch}: the matching position of itemset of query sequence; \textit{IIMatch}: the matching position of item in matching itemset of query sequence.}
	\If{\textit{IMatch} $==$ \textit{qs}.\textit{size}()}{
		output $s$;
	} 
	initializes $S_{temp}$ $\leftarrow$ $\varnothing$, $I_{temp}$ $\leftarrow$ $\varnothing$\;
	initializes \textit{newIMatch} $\leftarrow$ \textit{IMatch}, \textit{newIIMatch} $\leftarrow$ \textit{IIMatch}\;
	initializes \textit{qi} $\leftarrow$ null, $s\prime$ $\leftarrow$ $<$$>$, \textit{fn} $\leftarrow$ 0\;
	\For{$f$ $\in$ \textit{SE}}{
		\If{\textit{CMAP}(\textit{s}, $f$) $\textless$ \textit{minsup}}{
			continue;  \qquad(\textbf{Co-occurrence Pruning})
		}
		$s\prime$ $=$ $\leftarrow$ $s$ extend $f$ by $S$-Extention\;
		\If{\textit{sup}($s\prime$) $\ge$ \textit{minsup}}{
			$S_{temp}$ $=$ $S_{temp}$ $\cup$ $f$\;
		}
	}
	\For{$f$ $\in$ $S_{temp}$}{
		$s\prime$ $=$ $\leftarrow$ $s$ extend $f$ by $S$-Extention\;
		\textit{newIMatch}, \textit{newIIMatch} $=$ call \textbf{UpdateMatch}()\;
		\textit{fn} $=$ the result of comparing the bitmap of $f$ with the bitmap of \textit{qi}\;
		\If{\textit{fn} $\textless$ \textit{minsup}}{
			continue; \qquad(\textbf{USIP Strategy})
		}
		call \textbf{SEARCH}($s\prime$, $S_{temp}$, \{$e$ $\in$ $S_{temp}$ $\vert$ $e$ $\succ_{lex}$ $f$\}, \textit{qs}, \textit{minsup}, \textit{newIMatch}, \textit{newIIMatch}).
	}
	\For{$f$ $\in$ \textit{IE}}{
		\If{\textit{CMAP}(\textit{s}, $f$) $\textless$ \textit{minsup}}{
			continue;  \qquad(\textbf{Co-occurrence Pruning})
		}
		$s\prime$ $=$ $\leftarrow$ $s$ extend $f$ by $I$-Extention\;
		\If{\textit{sup}($s\prime$) $\ge$ \textit{minsup}}{
			$I_{temp}$ $=$ $I_{temp}$ $\cup$ $f$\;
		}
	}
	\For{$f$ $\in$ $I_{temp}$}{
		$s\prime$ $=$ $\leftarrow$ $s$ extend $f$ by $I$-Extention\;
		\textit{newIMatch}, \textit{newIIMatch} $=$ call \textbf{UpdateMatch}()\;
		\textit{fn} $=$ the result of comparing the bitmap of $f$ with the bitmap of \textit{qi}\;
		\If{\textit{fn} $\textless$ \textit{minsup}}{
			continue; \qquad(\textbf{UIIP Strategy})
		}
		call \textbf{SEARCH}($s\prime$, $S_{temp}$, \{$e$ $\in$ $I_{temp}$ $\vert$ $e$ $\succ_{lex}$ $f$\}, \textit{qs}, \textit{minsup}, \textit{newIMatch}, \textit{newIIMatch}).
	}
\end{algorithm}

The \textit{SEARCH} procedure is shown in Algorithm \ref{AL:SEARCH}. It first uses \textit{IMatch} to determine whether the current pattern exactly matches the query sequence, and outputs this frequent pattern if it does (Lines 1-3). Next, it initializes $S_{temp}$, $I_{temp}$, \textit{newIMatch}, \textit{newIIMatch}, \textit{qi}, $s\prime$, and \textit{fn} (Lines 4-6). $S_{temp}$ and $I_{temp}$ record which items can be extended by $S$-Extension or $I$-Extension respectively. \textit{newIMatch} and \textit{newIIMatch} record the new matching position of \textit{qi}. $s\prime$ is used to denote the extended sequence for $s$, and \textit{fn} is used by USIP and UIIP. And then, \textit{SEARCH} procedure finds out all items which are not skipped by co-occurrence pruning (Lines 7-10). \textit{SEARCH} uses these items to create $s\prime$ and judges whether $s\prime$ is frequent (Line 11). If $s\prime$ is frequent, extended item $f$ will be added into $S_{temp}$ (Lines 12-14). For all extended items in $S_{temp}$, \textit{SEARCH} will update \textit{newIMatch} and \textit{newIIMatch}, and obtain \textit{fn} (Lines 16-19). By utilizing USIP, the \textit{SEARCH} procedure will terminate some meaningless operations in $S$-Extension of $s$ (Lines 20-22). Subsequently, more target sequential patterns will be discovered recursively by the search procedure (Line 23). Similarly, the $I$-Extension process of $s$ is similar to that of $S$-Extension (Lines 25-42).

%


\section{Experiments} 
\label{sec:experiments}

In this section, we design two variants of the TaSPM algorithm to evaluate the effectiveness and efficiency of the proposed strategies. The first variant is called TaSPM$_{V1}$ which only uses the UTFP and a post-processing technique. Another is TaSPM$_{V2}$ that uses all proposed strategies including UTFP, UPIP, USIP, and UIIP. For TaSPM$_{V2}$, partial strategies are used along with the query sequence matching position. Therefore, TaSPM$_{V2}$ does not need to utilize this post-processing technique to save target sequences. To ensure that the number of obtained patterns is correct, we use the CM-SPAM algorithm \cite{fournier2014fast} as a comparison, which also makes use of the post-processing technique. TaSPM$_{V1}$ and TaSPM$_{V2}$ are compared with CM-SPAM in terms of running time, memory consumption, and number of bitmap intersections.

Our experimental equipment is a bare computer with 64-bit Windows 10 operating system. Its CPU is AMD Ryzen 5 3600 and has 8 GB RAM. All algorithms are implemented in Java language. The details of the experimental results are given below.

\subsection{Datasets for the Experiment}

We use six datasets for our experiments, and three of which are real datasets. These three real datasets are \textit{Bible}, \textit{Sign} and \textit{Accidents}. They are converted from books or real-life application records. Each itemset of these three real datasets has only one item, which means they are all single-item-based sequence datasets. These datasets are available for download from the open source platform SPMF\footnote{\url{http://www.philippe-fournier-viger.com/spmf/}}. \textit{Bible} is a moderately dense dataset and transformed from the book Bible. \textit{Sign} is a dense sign language dataset that has a few hundred items but many long sequences. \textit{Accidents} is an anonymized traffic accident dataset. It is so large that we chose a version containing 10k records in our experiments. As for synthetic datasets \textit{Syn10k}, \textit{Syn20k}, and \textit{Syn40k}, they are all multiple-items-based sequence datasets and generated from IBM data generator \cite{agrawal1995mining}. 

These synthetic datasets have many moderately long sequences, and can simulate some records of realistic applications very well. The details of all datasets using in experiments are shown in Table \ref{DATA}. For convenience, $\vert \mathcal{D} \vert$ is the size of dataset $\mathcal{D}$, $\vert \textit{I} \vert$ is the number of distinct items in $\mathcal{D}$, \textit{avg}($I$) is the average number of distinct items in a itemset of $\mathcal{D}$, \textit{avg}($S$) is the average length of sequences of $\mathcal{D}$, \textit{min}($S$) and \textit{max}($S$) are respectively the minimum and maximum size of sequences of $\mathcal{D}$.

\begin{table}[!h]
	\caption{Details of different experimental datasets}  
	\label{DATA}
	\centering
	\begin{tabular}{|c|c|c|c|c|c|c|}
		\hline
		\textbf{Dataset} & \textbf{$\vert \mathcal{D} \vert$} & \textbf{$\vert \textit{I} \vert$}  & \textbf{avg(I)} & \textbf{avg(S)} & \textbf{min(S)} & \textbf{max(S)} \\ \hline   \hline   
		Bible & 36369 & 13905 & 1.0 & 21.64 & 9 & 100  \\ \hline
		Sign	& 730 & 267	 & 1.0 & 51.99 & 18 & 94 \\ \hline
		Accidents10k	& 10000 & 310	 & 1.03 & 33.92 & 23 & 45 \\ \hline
		Syn10k	& 10000 & 7312 & 4.35 & 27.11 & 1 & 18  \\ \hline
		Syn20k	& 20000 & 7442	& 4.33  & 26.97 & 1 & 18 \\ \hline
		Syn40k & 40000 & 7537	 & 4.32 & 26.84 & 1 & 18 \\ \hline
	\end{tabular}
\end{table}

For six different datasets, we have selected six query sequences of size 3 for our experiments. They are $\textit{query}_\textit{Bible}$ = $<$(10), (46), (38)$>$, $\textit{query}_\textit{Sign}$ = $<$(17), (117), (144)$>$, $\textit{query}_\textit{Accidents}$ = $<$(16), (31), (43)$>$, $\textit{query}_\textit{Syn10k}$ = $<$(196), (8845, 9250), (4010)$>$, $\textit{query}_\textit{Syn20k}$ = $<$(1801), (842, 4616), (7752)$>$, and $\textit{query}_\textit{Syn40k}$ = $<$(8496), (5926, 6384), (9737)$>$. These sequences are randomly selected from the complete mining set. On the multiple-items-based synthetic dataset, the length of the query sequences is 4. This can ensure that the query sequences need to be extended through both $I$-Extension and $S$-Extension. Note that if a test point was run for more than 10000 minutes in an experiment, we consider this is unable to discover patterns. Correspondingly, its memory consumption will not be displayed.

\subsection{Efficiency Analysis}
\label{Efficiency Analysis}

\begin{figure}[!htbp]
	\centering
	\includegraphics[clip,scale=0.07]{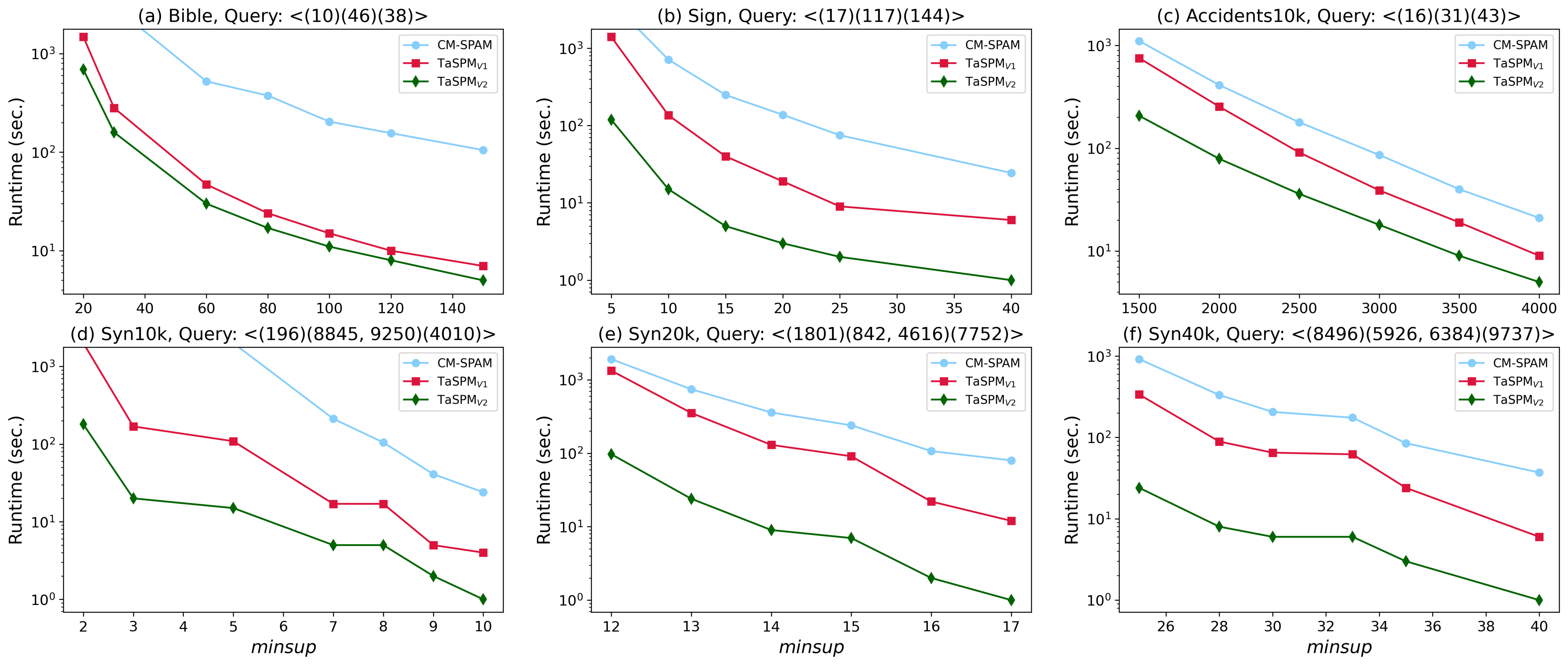}
	\caption{Runtime under various \textit{minsup}}
	\label{Runtime}
\end{figure}

In this subsection, we conduct a lot of experiments by adjusting \textit{minsup}. To better view the gap between algorithms, we present the logarithm of the runtime values. The experimental results are shown in Fig. \ref{Runtime}. From the results, it is clear that variants of the TaSPM algorithm can achieve better results compared to the baseline CM-SPAM algorithm. Even better, TaSPM$_{V2}$ can reduce the running time by more than one order of magnitude. 

Since TaSPM$_{V2}$ uses the most pruning strategies, it has the fastest running time. On the \textit{Bible} dataset, the runtime of TaSPM$_{V1}$ is 150\% to 200\% of the runtime of TaSPM$_{V2}$. This difference is not very significant compared to the other datasets. When \textit{minsup} is set to 40, baseline CM-SPAM can not mine any target pattern. However, even when \textit{minsup} is set to 20, both TaSPM$_{V1}$ and TaSPM$_{V2}$ are able to mine all patterns within 30 minutes. On the \textit{Sign} dataset, baseline CM-SPAM does not run when \textit{minsup} is 5. On the \textit{Accidents} dataset, the runtime difference between baseline CM-SPAM and TaSPM$_{V1}$ is not as large as it shows on the other dataset. This is because many sequences of the database contain the query sequence, and the database filters out few sequences. On the \textit{Syn10k} dataset, baseline CM-SPAM cannot mine any target sequential pattern when \textit{minsup} is set to 5, and TaSPM$_{V1}$ cannot mine any target sequential patterns when \textit{minsup} is set to 2. However, TaSPM$_{V2}$ is able to finish this task in about 200 seconds. Note that some of the runtime lines do not show a linear decrease, because when the \textit{minsup} is small, the change in \textit{minsup} does not make a significant change in the number of generated target sequential patterns. This also occurs in the other two synthetic datasets, but not obviously.

In addition, we also use the number of intersections to compare the efficiency of each algorithm, and the results are shown in Table \ref{time_inter}. Intersection is the main operation of the mining process, which largely affects the runtime. On the \textit{Bible} dataset, it is obvious that baseline CM-SPAM has 10 times more intersections than TaSPM$_{V1}$ and 20 times more intersections than TaSPM$_{V2}$. On the \textit{Sign} dataset, the number of intersections for CM-SPAM is tens of times more than that for TaSPM$_{V2}$. On the \textit{Accidents10k} dataset, the difference between CM-SPAM and TaSPM$_{V1}$ is more than double. On the synthetic dataset, for each algorithm, the number of intersections still differs greatly. TaSPM$_{V2}$ that uses the most pruning strategies always uses the least number of intersections, regardless of any dataset. This situation is more evident on some datasets.

\begin{table}[!htbp]
	\small
	\centering
	\caption{Number of intersection under various \textit{minsup}}
	\label{time_inter}
	\begin{tabular}{|c|c|p{1.3cm}<{\centering}|p{1.3cm}<{\centering}|p{1.3cm}<{\centering}|p{1.3cm}<{\centering}|p{1.3cm}<{\centering}|p{1.3cm}<{\centering}|}
		\hline \textbf{Dataset} &  \textbf{Algorithm} & $\mathbf{\textit{minsup}_{1}}$ & $\mathbf{\textit{minsup}_{2}}$ & $\mathbf{\textit{minsup}_{3}}$ & $\mathbf{\textit{minsup}_{4}}$ & $\mathbf{\textit{minsup}_{5}}$ & $\mathbf{\textit{minsup}_{6}}$\\
		\hline \hline
		\multirow{3}{*}{\makecell[c]{Bible}} 
		& {CM-SAPM} & {/} & {29,730,973} & {11,373,572} & {5,775,770} & {3,422,826} & {2,232,202}\\
		\cline{2-2}
		& {TaSPM$_{V1}$} & {10,696,148} & {3,796,616} & {1,205,362} & {536,414}  & {291,818} & {177,875}\\
		\cline{2-2}
		& {TaSPM$_{V2}$} & {4,613,035} & {1,714,605} & {550,128} &{249,246} &{136,309} & {83,073}\\
		\hline
		
		\multirow{3}{*}{\makecell[c]{Sign}} 
		& {CM-SPAM} & {/} & {326,469,305} & {93,050,762} & {38,131,279} & {19,044,929} & {10,709,256}\\
		\cline{2-2}
		& {TaSPM$_{V1}$} & {777,682,420} & {66,036,945} & {16,115,832} & {5,730,301} & {2,510,663} & {1,248,830}\\
		\cline{2-2}
		& {TaSPM$_{V2}$} & {61,293,211} & {5,464,723} & {1,426,235} &{544,759} & {261,804} & {144,437}\\
		\hline
		
		\multirow{3}{*}{\makecell[c]{Accidents10k}}
		& {CM-SPAM} & {5,411,453} & {1,678,640} & {590,381} & {243,176} & {111,509} & {54,163}\\
		\cline{2-2}
		& {TaSPM$_{V1}$} & {3,810,835} & {1,048,936} & {330,199} & {124,854} & {51,979} & {23,290}\\
		\cline{2-2}
		& {TaSPM$_{V2}$} & {807,694} & {257,933} & {94,368} &{39,531} & {17,953} & {8,428}\\
		\hline
		
		\multirow{3}{*}{\makecell[c]{Syn10k}}
		& {CM-SPAM} & {/} & {/} & {74,207,185} & {35,835,682} & {13,073,352} & {7,003,902}\\
		\cline{2-2}
		& {TaSPM$_{V1}$} & {54,487,018} & {33,554,406} & {5,336,006} & {5,111,782} & {1,556,454} & {1,080,294}\\
		\cline{2-2}
		& {TaSPM$_{V2}$} & {5,513,215} & {4,128,767} & {1,232,143} &{1,208,319} & {263,935} & {67,520}\\
		\hline
		
		\multirow{3}{*}{\makecell[c]{Syn20k}} 
		& {CM-SPAM} & {387,870,990} & {150,708,918} & {70,466,773} & {45,694,609} & {19,593,591} & {14,054,644}\\
		\cline{2-2}
		& {TaSPM$_{V1}$} & {266,338,270} & {66,846,686} & {24,004,574} & {16,920,542} & {4,081,633} & {2,240,481}\\
		\cline{2-2}
		& {TaSPM$_{V2}$} & {16,649,265} & {4,181,041} & {1,503,409} &{1,060,657} & {257,823} & {142,751}\\
		\hline
		
		\multirow{3}{*}{\makecell[c]{Syn40k}} 
		& {CM-SPAM}  & {100,042,878}& {27,263,239} & {19,970,108} & {13,631,370} & {7,414,568} & {2,587,062}\\
		\cline{2-2}
		& {TaSPM$_{V1}$} & {33,554,406} & {7,024,614} & {6,029,286} & {4,860,006} & {2,248,310} & {500,573}\\
		\cline{2-2}
		& {TaSPM$_{V2}$}  & {2,144,150}  & {486,038} & {423,830} &{350,750} & {176,126} & {51,871}\\
		\hline
		
		\hline
	\end{tabular}
\end{table}

\subsection{Memory Evaluation}
\label{Memory Evaluation}

\begin{figure}[!htbp]
	\centering
	\includegraphics[clip,scale=0.07]{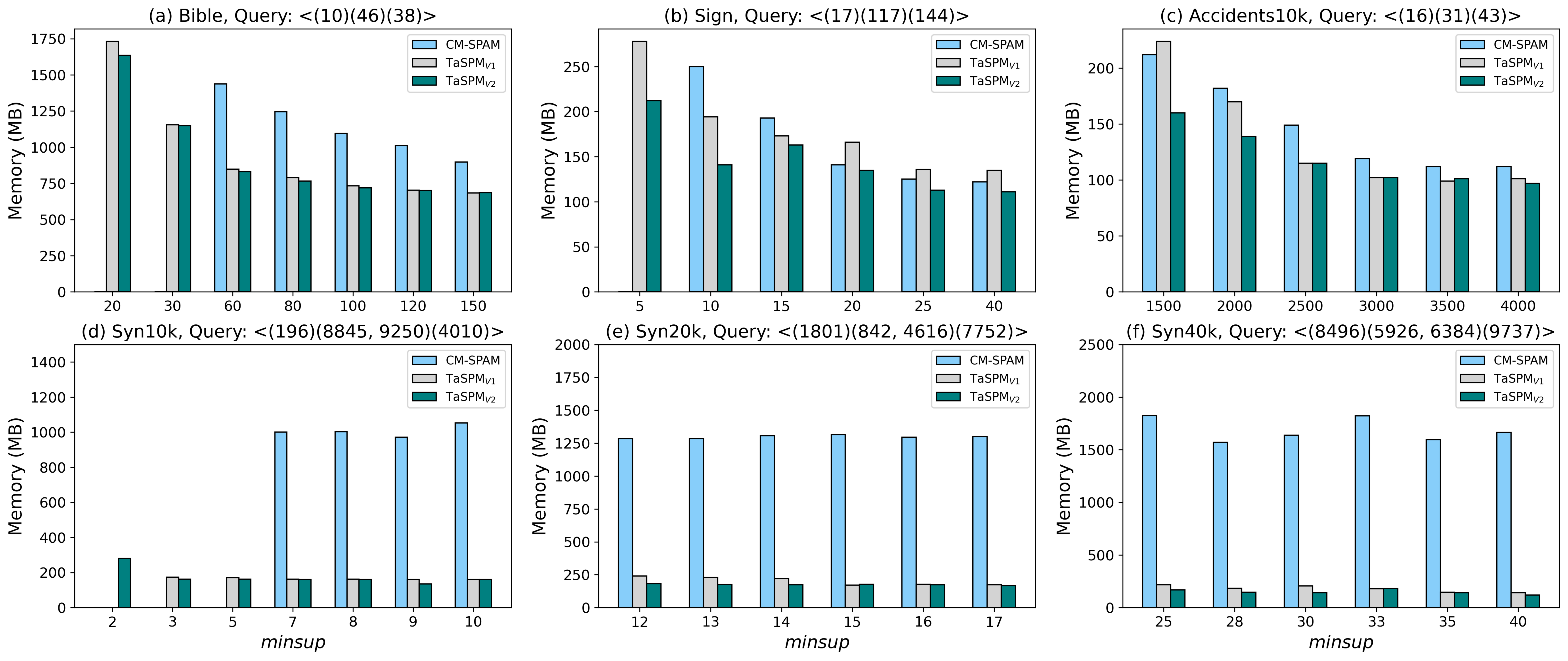}
	\caption{Memory usage under various \textit{minsup}}
	\label{memory}
\end{figure}

The results of each algorithm in terms of memory consumption are shown in the Fig. \ref{memory}. On the \textit{Bible} dataset, we can see that baseline algorithm CM-SPAM uses more memory compared to algorithms TaSPM$_{V1}$ and TaSPM$_{V2}$. This gap is slowly decreasing as the \textit{minsup} increases. 

Besides, at all test points, TaSPM$_{V1}$ uses a little more memory than TaSPM$_{V2}$. We can also see that the memory consumption shows a linear relationship with \textit{minsup} and plateaus at the end. On the \textit{Sign} dataset, TaSPM$_{V1}$ uses more memory when \textit{minsup} is greater than or equal to 20. As expected, TaSPM$_{V2}$ uses the least amount of memory. The memory consumption also decreases with the increase of \textit{minsup}. On the \textit{Accidents10k} dataset, the effect of \textit{minsup} is almost the same as on the \textit{Bible} dataset. However, the difference in memory consumption is not significant and there are a few more unstable test points. This is because the database filters few sequences. On the three synthetic datasets, we can see that  CM-SPAM uses several times more memory. This is because there are many records that do not contain the query sequence that are not filtered, increasing the memory consumption. In addition, the memory consumes in a fluctuating range of stable values. This is because the \textit{minsup} of the experiments is not changed vary much. Finally, we can still know that TaSPM$_{V2}$ is the most memory-efficient algorithm.

\subsection{Scalability}

\begin{figure}[ht]
	\centering
	\includegraphics[clip,scale=0.07]{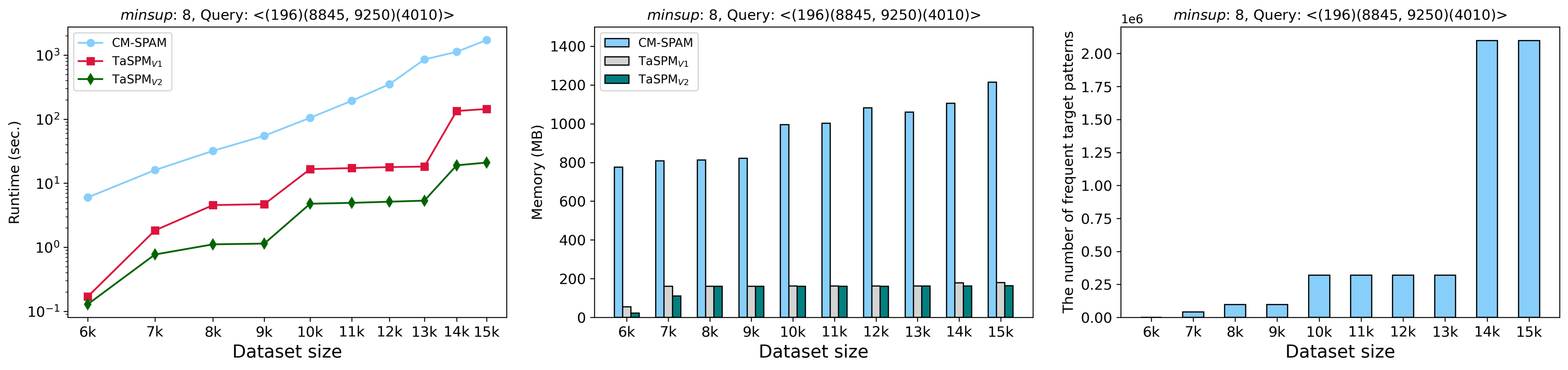}
	\caption{Scalability when \textit{minsup} = 8 and \textit{qs} = $<$(196), (8845, 9250), (4010)$>$}
	\label{scalability}
\end{figure}

In this subsection, we use 10 datasets to evaluate the scalability of each algorithm. We use \textit{qs} = $<$(196), (8845, 9250), (4010)$>$ as the query sequence and set \textit{minsup} to 8. The size of the dataset grows from 6000 sequences to 15000 sequences. We mainly evaluate the two aspects of runtime and memory consumption, and the experimental results are shown in the Fig. \ref{scalability}. It is clear that as the size of the dataset increases, the runtime of baseline CM-SPAM increases linearly. We can notice that TaSPM$_{V1}$ and TaSPM$_{V2}$ do not change much at some test points due to the increase in the size of the dataset, but the number of frequent target sequential patterns remains the same. This is because the runtime of CM-SPAM is affected by the size of the dataset, while the runtimes of TaSPM$_{V1}$ and TaSPM$_{V2}$ are affected by the number of target sequential patterns.

In terms of memory consumption, CM-SPAM uses several times more memory. The general trend of memory consumption for CM-SPAM also increases with dataset size. When the data setsize is between 8k and 15k, we can see that both TaSPM$_{V1}$ and TaSPM$_{V2}$ consume about 160MB of memory. In addition, the number of frequent target sequential patterns does not increase linearly. It remains the same or suddenly increase abruptly as the dataset size increases.

\subsection{Target sequence analysis}

\begin{figure}[ht]
	\centering
	\includegraphics[clip,scale=0.07]{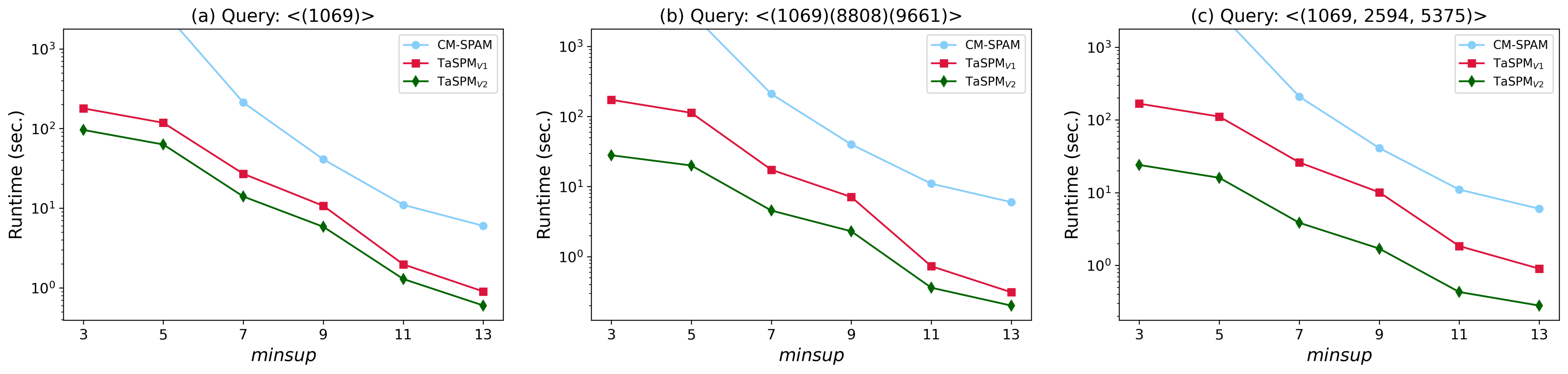}
	\caption{Runtime of different query sequences}
	\label{analysi_runtime}
\end{figure}

\begin{figure}[ht]
	\centering
	\includegraphics[clip,scale=0.07]{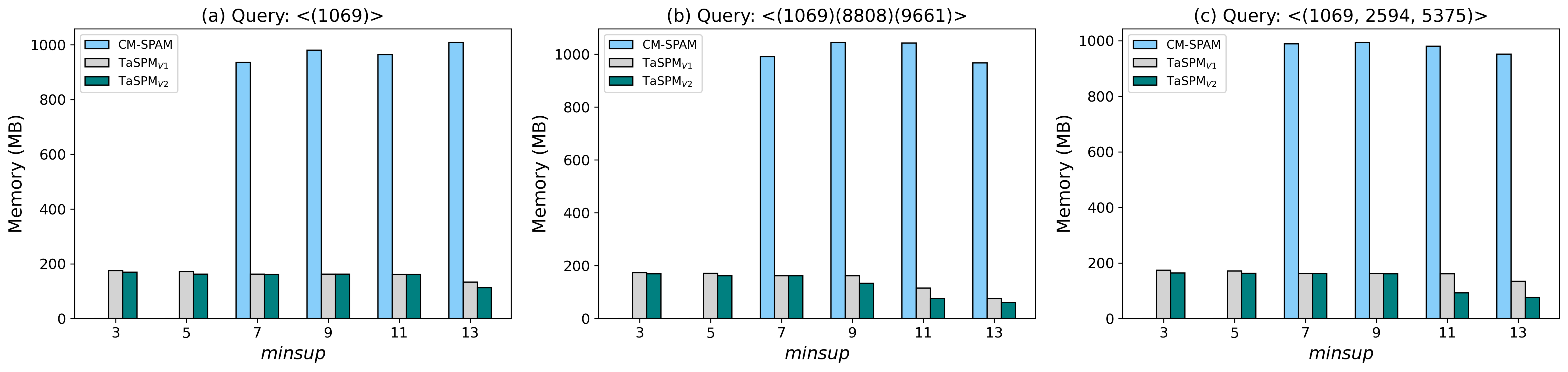}
	\caption{Memory of different query sequences}
	\label{analysi_memory}
\end{figure}

In subsection \ref{Efficiency Analysis}, the query sequence considers $I$-Extension and $S$-Extension, but does not consider some special query cases. To better evaluate the performance of TaSPM, we select different query sequences w.r.t. the target sequences to perform relevant experiments on the \textit{Syn10k} dataset. The query sequences are \textit{qs}$_{1}$ = $<$(1069)$>$, \textit{qs}$_{2}$ = $<$(1069), (8808), (9661)$>$, and \textit{qs}$_{3}$ = $<$(1069, 2594, 5375)$>$. Note that \textit{qs}$_{1}$ represents the case of querying only one item, \textit{qs}$_{2}$ represents the case of querying only sequences which extended through $S$-Extension, and \textit{qs}$_{3}$ represents the case of querying only sequences which extended through $I$-Extension. The experimental results are shown in Fig. \ref{analysi_runtime} and Fig. \ref{analysi_memory}, respectively.

From these results, we can see that under the three query sequences, CM-SPAM is still the least efficient. It fails to discover the interesting patterns when \textit{minsup} is reduced to 5. As the length of a query pattern increases, we can see that the running time of TaSPM$_{V2}$ on most datasets becomes less, but the running time of TaSPM$_{V1}$ does not change much. In terms of memory consumption, as analyzed in Section \ref{Memory Evaluation}, TaSPM$_{V2}$ uses the least amount of memory. The memory consumed by baseline CM-SPAM does not fluctuate much under different query sequences. In summary, we can learn that TaSPM$_{V2}$ is the best algorithm among these three variants for targeted sequential pattern mining.

\section{Conclusions}  
\label{sec:conclusion}

In this paper, to address a series of problems incurred by the current pattern mining and targeted pattern querying, we redefine two concepts and propose an efficient algorithm, called TaSPM, to quickly discover more focused results of target sequential patterns. Targeted mining makes more sense than mining all since the return results are more specific and make sense.  We use a post-processing technique to ensure that the mined patterns meet various requirements. In addition, we propose four pruning strategies to reduce meaningless intersections. Finally, a large number of experiments show that TaSPM can achieve better performance than baseline CM-SPAM for different datasets and different query sequences. In the future, we will propose different effective algorithms for specific applications based on the targeted pattern mining.

\section*{Acknowledgment}

This research was supported in part by the National Natural Science Foundation of China (Grant No. 62002136), Guangdong Basic and Applied Basic Research Foundation, Guangzhou Basic and Applied Basic Research Foundation (Grant No. 202102020277).


\bibliographystyle{ACM-Reference-Format}
\bibliography{main}


\begin{thebibliography}{47}


\ifx \showCODEN    \undefined \def \showCODEN     #1{\unskip}     \fi
\ifx \showDOI      \undefined \def \showDOI       #1{#1}\fi
\ifx \showISBNx    \undefined \def \showISBNx     #1{\unskip}     \fi
\ifx \showISBNxiii \undefined \def \showISBNxiii  #1{\unskip}     \fi
\ifx \showISSN     \undefined \def \showISSN      #1{\unskip}     \fi
\ifx \showLCCN     \undefined \def \showLCCN      #1{\unskip}     \fi
\ifx \shownote     \undefined \def \shownote      #1{#1}          \fi
\ifx \showarticletitle \undefined \def \showarticletitle #1{#1}   \fi
\ifx \showURL      \undefined \def \showURL       {\relax}        \fi
\providecommand\bibfield[2]{#2}
\providecommand\bibinfo[2]{#2}
\providecommand\natexlab[1]{#1}
\providecommand\showeprint[2][]{arXiv:#2}

\bibitem[\protect\citeauthoryear{Abeysinghe and Cui}{Abeysinghe and
  Cui}{2017}]%
        {abeysinghe2017query}
\bibfield{author}{\bibinfo{person}{Rashmie Abeysinghe} {and}
  \bibinfo{person}{Licong Cui}.} \bibinfo{year}{2017}\natexlab{}.
\newblock \showarticletitle{Query-constraint-based association rule mining from
  diverse clinical datasets in the national sleep research resource}. In
  \bibinfo{booktitle}{\emph{IEEE International Conference on Bioinformatics and
  Biomedicine}}. IEEE, \bibinfo{pages}{1238--1241}.
\newblock


\bibitem[\protect\citeauthoryear{Agrawal and Srikant}{Agrawal and
  Srikant}{1995}]%
        {agrawal1995mining}
\bibfield{author}{\bibinfo{person}{Rakesh Agrawal} {and}
  \bibinfo{person}{Ramakrishnan Srikant}.} \bibinfo{year}{1995}\natexlab{}.
\newblock \showarticletitle{Mining sequential patterns}. In
  \bibinfo{booktitle}{\emph{Proceedings of the 11th International Conference on
  Data Engineering}}. IEEE, \bibinfo{pages}{3--14}.
\newblock


\bibitem[\protect\citeauthoryear{Agrawal, Srikant, et~al\mbox{.}}{Agrawal
  et~al\mbox{.}}{1994}]%
        {agrawal1994fast}
\bibfield{author}{\bibinfo{person}{Rakesh Agrawal},
  \bibinfo{person}{Ramakrishnan Srikant}, {et~al\mbox{.}}}
  \bibinfo{year}{1994}\natexlab{}.
\newblock \showarticletitle{Fast algorithms for mining association rules}. In
  \bibinfo{booktitle}{\emph{Proceedings of the 20th International Conference on
  Very Large Data Bases}}. \bibinfo{pages}{487--499}.
\newblock


\bibitem[\protect\citeauthoryear{Ayres, Flannick, Gehrke, and Yiu}{Ayres
  et~al\mbox{.}}{2002}]%
        {ayres2002sequential}
\bibfield{author}{\bibinfo{person}{Jay Ayres}, \bibinfo{person}{Jason
  Flannick}, \bibinfo{person}{Johannes Gehrke}, {and} \bibinfo{person}{Tomi
  Yiu}.} \bibinfo{year}{2002}\natexlab{}.
\newblock \showarticletitle{Sequential pattern mining using a bitmap
  representation}. In \bibinfo{booktitle}{\emph{Proceedings of the 8th ACM
  SIGKDD International Conference on Knowledge Discovery and Data Mining}}.
  \bibinfo{pages}{429--435}.
\newblock


\bibitem[\protect\citeauthoryear{Chand, Thakkar, and Ganatra}{Chand
  et~al\mbox{.}}{2012}]%
        {chand2012target}
\bibfield{author}{\bibinfo{person}{Chetna Chand}, \bibinfo{person}{Amit
  Thakkar}, {and} \bibinfo{person}{Amit Ganatra}.}
  \bibinfo{year}{2012}\natexlab{}.
\newblock \showarticletitle{Target oriented sequential pattern mining using
  recency and monetary constraints}.
\newblock \bibinfo{journal}{\emph{International Journal of Computer
  Applications}} \bibinfo{volume}{45}, \bibinfo{number}{10}
  (\bibinfo{year}{2012}), \bibinfo{pages}{12--18}.
\newblock


\bibitem[\protect\citeauthoryear{Chiang, Wang, Lee, and Lin}{Chiang
  et~al\mbox{.}}{2003}]%
        {chiang2003goal}
\bibfield{author}{\bibinfo{person}{Ding~An Chiang}, \bibinfo{person}{Yi~Fan
  Wang}, \bibinfo{person}{Shao~Lun Lee}, {and} \bibinfo{person}{Cheng~Jung
  Lin}.} \bibinfo{year}{2003}\natexlab{}.
\newblock \showarticletitle{Goal-oriented sequential pattern for network
  banking churn analysis}.
\newblock \bibinfo{journal}{\emph{Expert Systems with Applications}}
  \bibinfo{volume}{25}, \bibinfo{number}{3} (\bibinfo{year}{2003}),
  \bibinfo{pages}{293--302}.
\newblock


\bibitem[\protect\citeauthoryear{Chueh}{Chueh}{2010}]%
        {chueh2010mining}
\bibfield{author}{\bibinfo{person}{Hao~En Chueh}.}
  \bibinfo{year}{2010}\natexlab{}.
\newblock \showarticletitle{Mining target-oriented sequential patterns with
  time-intervals}.
\newblock \bibinfo{journal}{\emph{arXiv preprint arXiv:1009.0929}}
  (\bibinfo{year}{2010}).
\newblock


\bibitem[\protect\citeauthoryear{Floratou, Tata, and Patel}{Floratou
  et~al\mbox{.}}{2011}]%
        {floratou2011efficient}
\bibfield{author}{\bibinfo{person}{Avrilia Floratou}, \bibinfo{person}{Sandeep
  Tata}, {and} \bibinfo{person}{Jignesh~M Patel}.}
  \bibinfo{year}{2011}\natexlab{}.
\newblock \showarticletitle{Efficient and accurate discovery of patterns in
  sequence data sets}.
\newblock \bibinfo{journal}{\emph{IEEE Transactions on Knowledge and Data
  Engineering}} \bibinfo{volume}{23}, \bibinfo{number}{8}
  (\bibinfo{year}{2011}), \bibinfo{pages}{1154--1168}.
\newblock


\bibitem[\protect\citeauthoryear{Fournier-Viger, Gomariz, Campos, and
  Thomas}{Fournier-Viger et~al\mbox{.}}{2014a}]%
        {fournier2014fast}
\bibfield{author}{\bibinfo{person}{Philippe Fournier-Viger},
  \bibinfo{person}{Antonio Gomariz}, \bibinfo{person}{Manuel Campos}, {and}
  \bibinfo{person}{Rincy Thomas}.} \bibinfo{year}{2014}\natexlab{a}.
\newblock \showarticletitle{Fast vertical mining of sequential patterns using
  co-occurrence information}. In \bibinfo{booktitle}{\emph{Pacific-Asia
  Conference on Knowledge Discovery and Data Mining}}. Springer,
  \bibinfo{pages}{40--52}.
\newblock


\bibitem[\protect\citeauthoryear{Fournier-Viger, Gomariz, Gueniche, Mwamikazi,
  and Thomas}{Fournier-Viger et~al\mbox{.}}{2013a}]%
        {fournier2013tks}
\bibfield{author}{\bibinfo{person}{Philippe Fournier-Viger},
  \bibinfo{person}{Antonio Gomariz}, \bibinfo{person}{Ted Gueniche},
  \bibinfo{person}{Esp{\'e}rance Mwamikazi}, {and} \bibinfo{person}{Rincy
  Thomas}.} \bibinfo{year}{2013}\natexlab{a}.
\newblock \showarticletitle{{TKS}: efficient mining of top-$k$ sequential
  patterns}. In \bibinfo{booktitle}{\emph{International International
  Conference on Advanced Data Mining and Applications}}. Springer,
  \bibinfo{pages}{109--120}.
\newblock


\bibitem[\protect\citeauthoryear{Fournier-Viger, Gomariz, {\v{S}}ebek, and
  Hlosta}{Fournier-Viger et~al\mbox{.}}{2014b}]%
        {fournier2014vgen}
\bibfield{author}{\bibinfo{person}{Philippe Fournier-Viger},
  \bibinfo{person}{Antonio Gomariz}, \bibinfo{person}{Michal {\v{S}}ebek},
  {and} \bibinfo{person}{Martin Hlosta}.} \bibinfo{year}{2014}\natexlab{b}.
\newblock \showarticletitle{{VGEN}: fast vertical mining of sequential
  generator patterns}. In \bibinfo{booktitle}{\emph{International Conference on
  Data Warehousing and Knowledge Discovery}}. Springer,
  \bibinfo{pages}{476--488}.
\newblock


\bibitem[\protect\citeauthoryear{Fournier-Viger, Gueniche, and
  Tseng}{Fournier-Viger et~al\mbox{.}}{2012}]%
        {fournier2012using}
\bibfield{author}{\bibinfo{person}{Philippe Fournier-Viger},
  \bibinfo{person}{Ted Gueniche}, {and} \bibinfo{person}{Vincent~S Tseng}.}
  \bibinfo{year}{2012}\natexlab{}.
\newblock \showarticletitle{Using partially-ordered sequential rules to
  generate more accurate sequence prediction}. In
  \bibinfo{booktitle}{\emph{International Conference on Advanced Data Mining
  and Applications}}. Springer, \bibinfo{pages}{431--442}.
\newblock


\bibitem[\protect\citeauthoryear{Fournier-Viger, Lin, Kiran, Koh, and
  Thomas}{Fournier-Viger et~al\mbox{.}}{2017}]%
        {fournier2017survey}
\bibfield{author}{\bibinfo{person}{Philippe Fournier-Viger},
  \bibinfo{person}{Jerry Chun~Wei Lin}, \bibinfo{person}{Rage~Uday Kiran},
  \bibinfo{person}{Yun~Sing Koh}, {and} \bibinfo{person}{Rincy Thomas}.}
  \bibinfo{year}{2017}\natexlab{}.
\newblock \showarticletitle{A survey of sequential pattern mining}.
\newblock \bibinfo{journal}{\emph{Data Science and Pattern Recognition}}
  \bibinfo{volume}{1}, \bibinfo{number}{1} (\bibinfo{year}{2017}),
  \bibinfo{pages}{54--77}.
\newblock


\bibitem[\protect\citeauthoryear{Fournier-Viger, Mwamikazi, Gueniche, and
  Faghihi}{Fournier-Viger et~al\mbox{.}}{2013b}]%
        {fournier2013meit}
\bibfield{author}{\bibinfo{person}{Philippe Fournier-Viger},
  \bibinfo{person}{Esp{\'e}rance Mwamikazi}, \bibinfo{person}{Ted Gueniche},
  {and} \bibinfo{person}{Usef Faghihi}.} \bibinfo{year}{2013}\natexlab{b}.
\newblock \showarticletitle{{MEIT}: Memory efficient itemset tree for targeted
  association rule mining}. In \bibinfo{booktitle}{\emph{International
  Conference on Advanced Data Mining and Applications}}. Springer,
  \bibinfo{pages}{95--106}.
\newblock


\bibitem[\protect\citeauthoryear{Fournier-Viger, Wu, Gomariz, and
  Tseng}{Fournier-Viger et~al\mbox{.}}{2014c}]%
        {fournier2014vmsp}
\bibfield{author}{\bibinfo{person}{Philippe Fournier-Viger},
  \bibinfo{person}{Cheng-Wei Wu}, \bibinfo{person}{Antonio Gomariz}, {and}
  \bibinfo{person}{Vincent~S Tseng}.} \bibinfo{year}{2014}\natexlab{c}.
\newblock \showarticletitle{{VMSP}: Efficient vertical mining of maximal
  sequential patterns}. In \bibinfo{booktitle}{\emph{Canadian Conference on
  Artificial Intelligence}}. Springer, \bibinfo{pages}{83--94}.
\newblock


\bibitem[\protect\citeauthoryear{Fournier-Viger, Wu, and Tseng}{Fournier-Viger
  et~al\mbox{.}}{2013c}]%
        {fournier2013mining}
\bibfield{author}{\bibinfo{person}{Philippe Fournier-Viger},
  \bibinfo{person}{Cheng-Wei Wu}, {and} \bibinfo{person}{Vincent~S Tseng}.}
  \bibinfo{year}{2013}\natexlab{c}.
\newblock \showarticletitle{Mining maximal sequential patterns without
  candidate maintenance}. In \bibinfo{booktitle}{\emph{International Conference
  on Advanced Data Mining and Applications}}. Springer,
  \bibinfo{pages}{169--180}.
\newblock


\bibitem[\protect\citeauthoryear{Fumarola, Lanotte, Ceci, and Malerba}{Fumarola
  et~al\mbox{.}}{2016}]%
        {fumarola2016clofast}
\bibfield{author}{\bibinfo{person}{Fabio Fumarola},
  \bibinfo{person}{Pasqua~Fabiana Lanotte}, \bibinfo{person}{Michelangelo
  Ceci}, {and} \bibinfo{person}{Donato Malerba}.}
  \bibinfo{year}{2016}\natexlab{}.
\newblock \showarticletitle{{CloFAST}: closed sequential pattern mining using
  sparse and vertical id-lists}.
\newblock \bibinfo{journal}{\emph{Knowledge and Information Systems}}
  \bibinfo{volume}{48}, \bibinfo{number}{2} (\bibinfo{year}{2016}),
  \bibinfo{pages}{429--463}.
\newblock


\bibitem[\protect\citeauthoryear{Gan, Du, Ding, Zhang, and Chao}{Gan
  et~al\mbox{.}}{2021a}]%
        {gan2021explainable}
\bibfield{author}{\bibinfo{person}{Wensheng Gan}, \bibinfo{person}{Zilin Du},
  \bibinfo{person}{Weiping Ding}, \bibinfo{person}{Chunkai Zhang}, {and}
  \bibinfo{person}{Han~Chieh Chao}.} \bibinfo{year}{2021}\natexlab{a}.
\newblock \showarticletitle{Explainable fuzzy utility mining on sequences}.
\newblock \bibinfo{journal}{\emph{IEEE Transactions on Fuzzy Systems}}
  \bibinfo{volume}{29}, \bibinfo{number}{12} (\bibinfo{year}{2021}),
  \bibinfo{pages}{3620 --3634}.
\newblock


\bibitem[\protect\citeauthoryear{Gan, Lin, Fournier-Viger, Chao, Tseng, and
  Yu}{Gan et~al\mbox{.}}{2021b}]%
        {gan2021survey}
\bibfield{author}{\bibinfo{person}{Wensheng Gan}, \bibinfo{person}{Jerry
  Chun~Wei Lin}, \bibinfo{person}{Philippe Fournier-Viger},
  \bibinfo{person}{Han~Chieh Chao}, \bibinfo{person}{Vincent~S Tseng}, {and}
  \bibinfo{person}{Philip~S Yu}.} \bibinfo{year}{2021}\natexlab{b}.
\newblock \showarticletitle{A survey of utility-oriented pattern mining}.
\newblock \bibinfo{journal}{\emph{IEEE Transactions on Knowledge and Data
  Engineering}} \bibinfo{volume}{33}, \bibinfo{number}{4}
  (\bibinfo{year}{2021}), \bibinfo{pages}{1306--1327}.
\newblock


\bibitem[\protect\citeauthoryear{Gan, Lin, Fournier-Viger, Chao, and Yu}{Gan
  et~al\mbox{.}}{2019}]%
        {gan2019survey}
\bibfield{author}{\bibinfo{person}{Wensheng Gan}, \bibinfo{person}{Jerry
  Chun~Wei Lin}, \bibinfo{person}{Philippe Fournier-Viger},
  \bibinfo{person}{Han~Chieh Chao}, {and} \bibinfo{person}{Philip~S Yu}.}
  \bibinfo{year}{2019}\natexlab{}.
\newblock \showarticletitle{A survey of parallel sequential pattern mining}.
\newblock \bibinfo{journal}{\emph{ACM Transactions on Knowledge Discovery from
  Data}} \bibinfo{volume}{13}, \bibinfo{number}{3} (\bibinfo{year}{2019}),
  \bibinfo{pages}{1--34}.
\newblock


\bibitem[\protect\citeauthoryear{Gan, Lin, Zhang, Chao, Fujita, and Yu}{Gan
  et~al\mbox{.}}{2020}]%
        {gan2020proum}
\bibfield{author}{\bibinfo{person}{Wensheng Gan}, \bibinfo{person}{Jerry
  Chun~Wei Lin}, \bibinfo{person}{Jiexiong Zhang}, \bibinfo{person}{Han~Chieh
  Chao}, \bibinfo{person}{Hamido Fujita}, {and} \bibinfo{person}{Philip~S Yu}.}
  \bibinfo{year}{2020}\natexlab{}.
\newblock \showarticletitle{{ProUM}: Projection-based utility mining on
  sequence data}.
\newblock \bibinfo{journal}{\emph{Information Sciences}}  \bibinfo{volume}{513}
  (\bibinfo{year}{2020}), \bibinfo{pages}{222--240}.
\newblock


\bibitem[\protect\citeauthoryear{Gan, Lin, Zhang, Yin, Fournier-Viger, Chao,
  and Yu}{Gan et~al\mbox{.}}{2021c}]%
        {gan2021utility}
\bibfield{author}{\bibinfo{person}{Wensheng Gan}, \bibinfo{person}{Jerry
  Chun~Wei Lin}, \bibinfo{person}{Jiexiong Zhang}, \bibinfo{person}{Hongzhi
  Yin}, \bibinfo{person}{Philippe Fournier-Viger}, \bibinfo{person}{Han~Chieh
  Chao}, {and} \bibinfo{person}{Philip~S Yu}.}
  \bibinfo{year}{2021}\natexlab{c}.
\newblock \showarticletitle{Utility mining across multi-dimensional sequences}.
\newblock \bibinfo{journal}{\emph{ACM Transactions on Knowledge Discovery from
  Data}} \bibinfo{volume}{15}, \bibinfo{number}{5} (\bibinfo{year}{2021}),
  \bibinfo{pages}{1--24}.
\newblock


\bibitem[\protect\citeauthoryear{Gao, Wang, He, and Zhou}{Gao
  et~al\mbox{.}}{2008}]%
        {gao2008efficient}
\bibfield{author}{\bibinfo{person}{Chuancong Gao}, \bibinfo{person}{Jianyong
  Wang}, \bibinfo{person}{Yukai He}, {and} \bibinfo{person}{Lizhu Zhou}.}
  \bibinfo{year}{2008}\natexlab{}.
\newblock \showarticletitle{Efficient mining of frequent sequence generators}.
  In \bibinfo{booktitle}{\emph{Proceedings of the 17th International Conference
  on World Wide Web}}. \bibinfo{pages}{1051--1052}.
\newblock


\bibitem[\protect\citeauthoryear{Garc{\'\i}a-Hern{\'a}ndez,
  Mart{\'\i}nez-Trinidad, and Carrasco-Ochoa}{Garc{\'\i}a-Hern{\'a}ndez
  et~al\mbox{.}}{2006}]%
        {garcia2006new}
\bibfield{author}{\bibinfo{person}{Ren{\'e}~Arnulfo Garc{\'\i}a-Hern{\'a}ndez},
  \bibinfo{person}{Jos{\'e}~Francisco Mart{\'\i}nez-Trinidad}, {and}
  \bibinfo{person}{Jes{\'u}s~Ariel Carrasco-Ochoa}.}
  \bibinfo{year}{2006}\natexlab{}.
\newblock \showarticletitle{A new algorithm for fast discovery of maximal
  sequential patterns in a document collection}. In
  \bibinfo{booktitle}{\emph{International Conference on Intelligent Text
  Processing and Computational Linguistics}}. Springer,
  \bibinfo{pages}{514--523}.
\newblock


\bibitem[\protect\citeauthoryear{Gomariz, Campos, Marin, and Goethals}{Gomariz
  et~al\mbox{.}}{2013}]%
        {gomariz2013clasp}
\bibfield{author}{\bibinfo{person}{Antonio Gomariz}, \bibinfo{person}{Manuel
  Campos}, \bibinfo{person}{Roque Marin}, {and} \bibinfo{person}{Bart
  Goethals}.} \bibinfo{year}{2013}\natexlab{}.
\newblock \showarticletitle{{ClaSP}: An efficient algorithm for mining frequent
  closed sequences}. In \bibinfo{booktitle}{\emph{Pacific-Asia Conference on
  Knowledge Discovery and Data Mining}}. Springer, \bibinfo{pages}{50--61}.
\newblock


\bibitem[\protect\citeauthoryear{Guan, Chang, Wang, and Zhou}{Guan
  et~al\mbox{.}}{2005}]%
        {guan2005mining}
\bibfield{author}{\bibinfo{person}{En~Zheng Guan}, \bibinfo{person}{Xiao~Yu
  Chang}, \bibinfo{person}{Zhe Wang}, {and} \bibinfo{person}{Chun~Guang Zhou}.}
  \bibinfo{year}{2005}\natexlab{}.
\newblock \showarticletitle{Mining maximal sequential patterns}. In
  \bibinfo{booktitle}{\emph{International Conference on Neural Networks and
  Brain}}, Vol.~\bibinfo{volume}{1}. IEEE, \bibinfo{pages}{525--528}.
\newblock


\bibitem[\protect\citeauthoryear{Han, Pei, Mortazavi-Asl, Pinto, Chen, Dayal,
  and Hsu}{Han et~al\mbox{.}}{2001}]%
        {han2001prefixspan}
\bibfield{author}{\bibinfo{person}{Jiawei Han}, \bibinfo{person}{Jian Pei},
  \bibinfo{person}{Behzad Mortazavi-Asl}, \bibinfo{person}{Helen Pinto},
  \bibinfo{person}{Qiming Chen}, \bibinfo{person}{Umeshwar Dayal}, {and}
  \bibinfo{person}{Meichun Hsu}.} \bibinfo{year}{2001}\natexlab{}.
\newblock \showarticletitle{{PrefixSpan}: Mining sequential patterns
  efficiently by prefix-projected pattern growth}. In
  \bibinfo{booktitle}{\emph{Proceedings of the 17th International Conference on
  Data Engineering}}. IEEE, \bibinfo{pages}{215--224}.
\newblock


\bibitem[\protect\citeauthoryear{Kubat, Hafez, Raghavan, Lekkala, and
  Chen}{Kubat et~al\mbox{.}}{2003}]%
        {kubat2003itemset}
\bibfield{author}{\bibinfo{person}{Miroslav Kubat}, \bibinfo{person}{Aladdin
  Hafez}, \bibinfo{person}{Vijay~V Raghavan}, \bibinfo{person}{Jayakrishna~R
  Lekkala}, {and} \bibinfo{person}{Wei~Kian Chen}.}
  \bibinfo{year}{2003}\natexlab{}.
\newblock \showarticletitle{Itemset trees for targeted association querying}.
\newblock \bibinfo{journal}{\emph{IEEE Transactions on Knowledge and Data
  Engineering}} \bibinfo{volume}{15}, \bibinfo{number}{6}
  (\bibinfo{year}{2003}), \bibinfo{pages}{1522--1534}.
\newblock


\bibitem[\protect\citeauthoryear{Lam, M{\"o}rchen, Fradkin, and Calders}{Lam
  et~al\mbox{.}}{2014}]%
        {lam2014mining}
\bibfield{author}{\bibinfo{person}{Hoang~Thanh Lam}, \bibinfo{person}{Fabian
  M{\"o}rchen}, \bibinfo{person}{Dmitriy Fradkin}, {and} \bibinfo{person}{Toon
  Calders}.} \bibinfo{year}{2014}\natexlab{}.
\newblock \showarticletitle{Mining compressing sequential patterns}.
\newblock \bibinfo{journal}{\emph{Statistical Analysis and Data Mining}}
  \bibinfo{volume}{7}, \bibinfo{number}{1} (\bibinfo{year}{2014}),
  \bibinfo{pages}{34--52}.
\newblock


\bibitem[\protect\citeauthoryear{Lewis, Benton, Bourrie, and Lavergne}{Lewis
  et~al\mbox{.}}{2019}]%
        {lewis2019enhancing}
\bibfield{author}{\bibinfo{person}{Jay Lewis}, \bibinfo{person}{Ryan~G Benton},
  \bibinfo{person}{David Bourrie}, {and} \bibinfo{person}{Jennifer Lavergne}.}
  \bibinfo{year}{2019}\natexlab{}.
\newblock \showarticletitle{Enhancing itemset tree rules and performance}. In
  \bibinfo{booktitle}{\emph{IEEE International Conference on Big Data}}. IEEE,
  \bibinfo{pages}{1143--1150}.
\newblock


\bibitem[\protect\citeauthoryear{Miao, Wan, Gan, Sun, and Chen}{Miao
  et~al\mbox{.}}{2021}]%
        {miao2021targetum}
\bibfield{author}{\bibinfo{person}{Jinbao Miao}, \bibinfo{person}{Shicheng
  Wan}, \bibinfo{person}{Wensheng Gan}, \bibinfo{person}{Jiayi Sun}, {and}
  \bibinfo{person}{Jiahui Chen}.} \bibinfo{year}{2021}\natexlab{}.
\newblock \showarticletitle{Targeted high-utility itemset querying}. In
  \bibinfo{booktitle}{\emph{IEEE International Conference on Big Data}}. IEEE,
  \bibinfo{pages}{5534--5543}.
\newblock


\bibitem[\protect\citeauthoryear{Mooney and Roddick}{Mooney and
  Roddick}{2013}]%
        {mooney2013sequential}
\bibfield{author}{\bibinfo{person}{Carl~H Mooney} {and} \bibinfo{person}{John~F
  Roddick}.} \bibinfo{year}{2013}\natexlab{}.
\newblock \showarticletitle{Sequential pattern mining-approaches and
  algorithms}.
\newblock \bibinfo{journal}{\emph{Comput. Surveys}} \bibinfo{volume}{45},
  \bibinfo{number}{2} (\bibinfo{year}{2013}), \bibinfo{pages}{1--39}.
\newblock


\bibitem[\protect\citeauthoryear{Petitjean, Li, Tatti, and Webb}{Petitjean
  et~al\mbox{.}}{2016}]%
        {petitjean2016skopus}
\bibfield{author}{\bibinfo{person}{Fran{\c{c}}ois Petitjean},
  \bibinfo{person}{Tao Li}, \bibinfo{person}{Nikolaj Tatti}, {and}
  \bibinfo{person}{Geoffrey~I Webb}.} \bibinfo{year}{2016}\natexlab{}.
\newblock \showarticletitle{{SkOPUS}: Mining top-$k$ sequential patterns under
  leverage}.
\newblock \bibinfo{journal}{\emph{Data Mining and Knowledge Discovery}}
  \bibinfo{volume}{30}, \bibinfo{number}{5} (\bibinfo{year}{2016}),
  \bibinfo{pages}{1086--1111}.
\newblock


\bibitem[\protect\citeauthoryear{Pinto, Han, Pei, Wang, Chen, and Dayal}{Pinto
  et~al\mbox{.}}{2001}]%
        {pinto2001multi}
\bibfield{author}{\bibinfo{person}{Helen Pinto}, \bibinfo{person}{Jiawei Han},
  \bibinfo{person}{Jian Pei}, \bibinfo{person}{Ke Wang},
  \bibinfo{person}{Qiming Chen}, {and} \bibinfo{person}{Umeshwar Dayal}.}
  \bibinfo{year}{2001}\natexlab{}.
\newblock \showarticletitle{Multi-dimensional sequential pattern mining}. In
  \bibinfo{booktitle}{\emph{Proceedings of the 10th International Conference on
  Information and Knowledge Management}}. \bibinfo{pages}{81--88}.
\newblock


\bibitem[\protect\citeauthoryear{Salvemini, Fumarola, Malerba, and
  Han}{Salvemini et~al\mbox{.}}{2011}]%
        {salvemini2011fast}
\bibfield{author}{\bibinfo{person}{Eliana Salvemini}, \bibinfo{person}{Fabio
  Fumarola}, \bibinfo{person}{Donato Malerba}, {and} \bibinfo{person}{Jiawei
  Han}.} \bibinfo{year}{2011}\natexlab{}.
\newblock \showarticletitle{Fast sequence mining based on sparse id-lists}. In
  \bibinfo{booktitle}{\emph{International Symposium on Methodologies for
  Intelligent Systems}}. Springer, \bibinfo{pages}{316--325}.
\newblock


\bibitem[\protect\citeauthoryear{Shabtay, Fournier-Viger, Yaari, and
  Dattner}{Shabtay et~al\mbox{.}}{2021}]%
        {shabtay2021guided}
\bibfield{author}{\bibinfo{person}{Lior Shabtay}, \bibinfo{person}{Philippe
  Fournier-Viger}, \bibinfo{person}{Rami Yaari}, {and} \bibinfo{person}{Itai
  Dattner}.} \bibinfo{year}{2021}\natexlab{}.
\newblock \showarticletitle{A guided {FP-Growth} algorithm for mining
  multitude-targeted item-sets and class association rules in imbalanced data}.
\newblock \bibinfo{journal}{\emph{Information Sciences}}  \bibinfo{volume}{553}
  (\bibinfo{year}{2021}), \bibinfo{pages}{353--375}.
\newblock


\bibitem[\protect\citeauthoryear{Songram and Boonjing}{Songram and
  Boonjing}{2008}]%
        {songram2008closed}
\bibfield{author}{\bibinfo{person}{Panida Songram} {and} \bibinfo{person}{Veera
  Boonjing}.} \bibinfo{year}{2008}\natexlab{}.
\newblock \showarticletitle{Closed multidimensional sequential pattern mining}.
\newblock \bibinfo{journal}{\emph{International Journal of Knowledge Management
  Studies}} \bibinfo{volume}{2}, \bibinfo{number}{4} (\bibinfo{year}{2008}),
  \bibinfo{pages}{460--479}.
\newblock


\bibitem[\protect\citeauthoryear{Srikant and Agrawal}{Srikant and
  Agrawal}{1996}]%
        {srikant1996mining}
\bibfield{author}{\bibinfo{person}{Ramakrishnan Srikant} {and}
  \bibinfo{person}{Rakesh Agrawal}.} \bibinfo{year}{1996}\natexlab{}.
\newblock \showarticletitle{Mining sequential patterns: generalizations and
  performance improvements}. In \bibinfo{booktitle}{\emph{Proceeding of the
  International Conference on Extending Database Technology}}. Springer,
  \bibinfo{pages}{1--17}.
\newblock


\bibitem[\protect\citeauthoryear{Tzvetkov, Yan, and Han}{Tzvetkov
  et~al\mbox{.}}{2005}]%
        {tzvetkov2005tsp}
\bibfield{author}{\bibinfo{person}{Petre Tzvetkov}, \bibinfo{person}{Xifeng
  Yan}, {and} \bibinfo{person}{Jiawei Han}.} \bibinfo{year}{2005}\natexlab{}.
\newblock \showarticletitle{{TSP}: Mining top-$k$ closed sequential patterns}.
\newblock \bibinfo{journal}{\emph{Knowledge and Information Systems}}
  \bibinfo{volume}{7}, \bibinfo{number}{4} (\bibinfo{year}{2005}),
  \bibinfo{pages}{438--457}.
\newblock


\bibitem[\protect\citeauthoryear{Wang, Han, and Li}{Wang et~al\mbox{.}}{2007}]%
        {wang2007frequent}
\bibfield{author}{\bibinfo{person}{Jianyong Wang}, \bibinfo{person}{Jiawei
  Han}, {and} \bibinfo{person}{Chun Li}.} \bibinfo{year}{2007}\natexlab{}.
\newblock \showarticletitle{Frequent closed sequence mining without candidate
  maintenance}.
\newblock \bibinfo{journal}{\emph{IEEE Transactions on Knowledge and Data
  Engineering}} \bibinfo{volume}{19}, \bibinfo{number}{8}
  (\bibinfo{year}{2007}), \bibinfo{pages}{1042--1056}.
\newblock


\bibitem[\protect\citeauthoryear{Wu, Tong, Zhu, and Wu}{Wu
  et~al\mbox{.}}{2017}]%
        {wu2017nosep}
\bibfield{author}{\bibinfo{person}{Youxi Wu}, \bibinfo{person}{Yao Tong},
  \bibinfo{person}{Xingquan Zhu}, {and} \bibinfo{person}{Xindong Wu}.}
  \bibinfo{year}{2017}\natexlab{}.
\newblock \showarticletitle{{NOSEP}: Nonoverlapping sequence pattern mining
  with gap constraints}.
\newblock \bibinfo{journal}{\emph{IEEE Transactions on Cybernetics}}
  \bibinfo{volume}{48}, \bibinfo{number}{10} (\bibinfo{year}{2017}),
  \bibinfo{pages}{2809--2822}.
\newblock


\bibitem[\protect\citeauthoryear{Wu, Zhu, Li, Guo, and Wu}{Wu
  et~al\mbox{.}}{2020}]%
        {wu2020netncsp}
\bibfield{author}{\bibinfo{person}{Youxi Wu}, \bibinfo{person}{Changrui Zhu},
  \bibinfo{person}{Yan Li}, \bibinfo{person}{Lei Guo}, {and}
  \bibinfo{person}{Xindong Wu}.} \bibinfo{year}{2020}\natexlab{}.
\newblock \showarticletitle{{NetNCSP}: Nonoverlapping closed sequential pattern
  mining}.
\newblock \bibinfo{journal}{\emph{Knowledge-Based Systems}}
  \bibinfo{volume}{196} (\bibinfo{year}{2020}), \bibinfo{pages}{105812}.
\newblock


\bibitem[\protect\citeauthoryear{Yan, Han, and Afshar}{Yan
  et~al\mbox{.}}{2003}]%
        {yan2003clospan}
\bibfield{author}{\bibinfo{person}{Xifeng Yan}, \bibinfo{person}{Jiawei Han},
  {and} \bibinfo{person}{Ramin Afshar}.} \bibinfo{year}{2003}\natexlab{}.
\newblock \showarticletitle{{CloSpan}: Mining closed sequential patterns in
  large datasets}. In \bibinfo{booktitle}{\emph{Proceedings of the 3rd SIAM
  International Conference on Data Mining}}. SIAM, \bibinfo{pages}{166--177}.
\newblock


\bibitem[\protect\citeauthoryear{Yang and M}{Yang and M}{2005}]%
        {yang2005improved}
\bibfield{author}{\bibinfo{person}{Zhenglu Yang} {and}
  \bibinfo{person}{Kitsuregawa M}.} \bibinfo{year}{2005}\natexlab{}.
\newblock \showarticletitle{{LAPIN-SPAM}: An improved algorithm for mining
  sequential pattern}. In \bibinfo{booktitle}{\emph{International Conference on
  Data Engineering Workshops}}. \bibinfo{pages}{1222--1225}.
\newblock


\bibitem[\protect\citeauthoryear{Yang, Wang, and Kitsuregawa}{Yang
  et~al\mbox{.}}{2007}]%
        {yang2007lapin}
\bibfield{author}{\bibinfo{person}{Zhenglu Yang}, \bibinfo{person}{Yitong
  Wang}, {and} \bibinfo{person}{Masaru Kitsuregawa}.}
  \bibinfo{year}{2007}\natexlab{}.
\newblock \showarticletitle{{LAPIN}: effective sequential pattern mining
  algorithms by last position induction for dense databases}. In
  \bibinfo{booktitle}{\emph{International Conference on Database Systems for
  Advanced Applications}}. Springer, \bibinfo{pages}{1020--1023}.
\newblock


\bibitem[\protect\citeauthoryear{Zaki}{Zaki}{2001}]%
        {zaki2001spade}
\bibfield{author}{\bibinfo{person}{Mohammed~J Zaki}.}
  \bibinfo{year}{2001}\natexlab{}.
\newblock \showarticletitle{{SPADE}: An efficient algorithm for mining frequent
  sequences}.
\newblock \bibinfo{journal}{\emph{Machine Learning}} \bibinfo{volume}{42},
  \bibinfo{number}{1-2} (\bibinfo{year}{2001}), \bibinfo{pages}{31--60}.
\newblock


\bibitem[\protect\citeauthoryear{Zhang, Du, Dai, Gan, Weng, and Yu}{Zhang
  et~al\mbox{.}}{2021}]%
        {zhang2021tusq}
\bibfield{author}{\bibinfo{person}{Chunkai Zhang}, \bibinfo{person}{Zilin Du},
  \bibinfo{person}{Quanjian Dai}, \bibinfo{person}{Wensheng Gan},
  \bibinfo{person}{Jian Weng}, {and} \bibinfo{person}{Philip~S Yu}.}
  \bibinfo{year}{2021}\natexlab{}.
\newblock \showarticletitle{{TUSQ}: Targeted high-utility sequence querying}.
\newblock \bibinfo{journal}{\emph{IEEE Transactions on Big Data.
  arXiv:2103.16615}} (\bibinfo{year}{2021}), \bibinfo{pages}{1--14}.
\newblock


\end{thebibliography}


\end{document}